\documentclass[prd,aps,showpacs,nofootinbib,tightenlines]{revtex4}
\usepackage{mathrsfs}
\usepackage{amsmath}
\usepackage{amssymb}
\usepackage{epsfig}
\usepackage{graphicx}
\usepackage{booktabs}
\usepackage{multirow}
\usepackage{subfigure}
\usepackage[section]{placeins}
\usepackage{float}
\usepackage{slashed}
\usepackage{color}
\begin{document}
\newcommand{\psl}{ p \hspace{-1.8truemm}/ }
\newcommand{\nsl}{ n \hspace{-2.2truemm}/ }
\newcommand{\vsl}{ v \hspace{-2.2truemm}/ }
\newcommand{\epsl}{\epsilon \hspace{-1.8truemm}/\,  }

\title{Investigating the color-suppressed decays $\Lambda_b\rightarrow \Lambda  \psi$ in the
perturbative QCD approach }
\author{Zhou Rui$^1$}\email[Corresponding  author: ]{jindui1127@126.com}
\author{Chao-Qi Zhang$^1$}
\author{Jia-Ming Li$^1$}
\author{Meng-Kun Jia$^1$}
\affiliation{$^1$College of Sciences, North China University of Science and Technology,
                          Tangshan 063009,  China}
\date{\today}
\begin{abstract}

The nonleptonic two-body $\Lambda_b\rightarrow \Lambda\psi $ decays with $\psi=J/\psi$ or $\psi(2S)$
are investigated based on the perturbative QCD approach.
These are color-suppressed processes in which the nonfactorizable contributions are confirmed to be dominant.
Angular momentum conservation allows us to describe the concerned  decays by four independent complex helicity amplitudes.
It is observed that the negative-helicity states for the $\Lambda$ baryon are preferred
as expected in the left-handed nature of the charged-current interaction.
The obtained results for the helicity amplitudes  are used to compute the branching ratios and various observable parameters,
which are then compared to the existing theoretical predictions and experimental data.
In particular, we predict the ratio $\mathcal{R}=\frac{\mathcal{B}(\Lambda_b\rightarrow\Lambda \psi(2S))}{\mathcal{B}(\Lambda_b\rightarrow\Lambda J/\psi)}=0.47^{+0.02}_{-0.00}$ in comparison with $0.508\pm 0.023$ from the Particle Data Group at the level of 1 standard deviation.
We also briefly explore the long-distance contributions to the semileptonic $\Lambda_b \rightarrow \Lambda l^+l^-$ decays
in the kinematic regions where the dilepton invariant masses are around the $J/\psi$  and  $\psi(2S)$  resonances.

\end{abstract}



\maketitle

\section{Introduction}
Exclusive decays of $b$-flavored hadrons into charmonia, governed by the weak $b\rightarrow s c \bar c$ transition,
provide valuable insight into the dynamics of strong interactions in the heavy hadronic decays.
The charmonium mode belongs to the color-suppressed category~\cite{Neubert:2001sj},
which receives large nonfactorizable contributions  and poses a challenge for the factorization ansatz.
Such processes have been the subject of theoretical and experimental interest in bottom meson decays,
such as $B\rightarrow J/\psi K$~\cite{CDF:1995izg,Belle:2002oex,BaBar:2004htr,prd71114008,Cheng:2000kt}.
In the baryon sector, the typical process is the $\Lambda_b \rightarrow \Lambda J/\psi$ decay,
where the  QCD dynamics are more complicated by the presence of extra spectator quarks.
As $\Lambda_b$ has nonzero spin,
this mode is a useful environment in which to study the helicity
structure of the underlying Hamiltonian~\cite{Buchalla:1995vs,Mannel:1997pc,Hiller:2007ur}.

The $b$-hadron decays into $J/\psi$ are experimentally convenient
because the subsequent decay $J/\psi\rightarrow \mu^+\mu^-$ has particularly distinctive signatures.
The $\Lambda_b \rightarrow \Lambda J/\psi$ mode was first observed by the UA1 Collaboration
at the CERN proton-antiproton ($p\bar p$) collider~\cite{UA1:1991vse},
followed by extensive studies at the Fermilab Tevatron by the CDF~\cite{CDF:1992lrw,CDF:1996rvy,CDF:2006eul}
and D0~\cite{D0:2007giz,D0:2004quf,prd84031102,prd85112003} Collaborations.
However, its absolute branching ratio has not yet been determined,
since the experimental knowledge of the fraction of $b$ quarks which hadronize to $\Lambda_b$ baryons is currently limited.
The current Particle Data Group (PDG) presents an average value~\cite{pdg2020}
\begin{eqnarray}\label{eq:bbb}
f(b\rightarrow\Lambda_b)\times \mathcal{B}(\Lambda_b \rightarrow \Lambda J/\psi)=(5.8\pm0.8)\times10^{-5},
\end{eqnarray}
where $f(b\rightarrow\Lambda_b)$ describes the probability that a $b$ quark fragments into a $\Lambda_b$ baryon.
Another salient feature of the decay is the wealth of information carried by angular observables
in terms of angular asymmetries that can be exploited to probe new physics beyond the standard model.
An angular analysis of the decay  $\Lambda_b \rightarrow \Lambda J/\psi$ was first done by the LHCb Collaboration~\cite{plb72427},
where the $\Lambda_b$'s are produced in $pp$ collisions  at the Large Hadron Collider (LHC).
Subsequently, a similar analysis was also performed by the ATLAS~\cite{prd89092009} and CMS~\cite{CMS:2016iaf,prd97072010} Collaborations.
Some interesting observables, such as the helicity amplitudes, production polarization,
the parity-violating parameter and other asymmetry parameters, are now available.
A latest analysis was conducted by the LHCb Collaboration~\cite{jhep062020110},
in which the polarization of $\Lambda_b$ baryons is measured for the first time at  $\sqrt{s}=13$ TeV.
All of these measurements  show that the production polarization  of $\Lambda_b$ is consistent with zero.

As a tremendous amount of beauty baryons is produced at the LHC,
numerous decays of the $\Lambda_b$ baryon to excited charmonium states have been
 observed~\cite{LHCb:2016hey,LHCb:2017zzt,LHCb:2018qxv,LHCb:2019imv}.
Among these, the decay mode  $\Lambda_b\rightarrow \Lambda \psi(2S)$  is of particular interest because
it is the radial excited  partner of  $\Lambda_b\rightarrow \Lambda J/\psi$ with the same topology.
It could  provide  additional and complementary phenomenological information on the QCD dynamics of the charmonium $\Lambda_b$ decays.
The first observation of $\Lambda_b\rightarrow \Lambda \psi(2S)$  was made by the ATLAS Collaboration~\cite{plb75163}.
Later, the reaction was also observed by the LHCb Collaboration~\cite{jhep032019126}.
Similarly, no absolute branching ratio is measured for this decay 
and only ratios to other reactions are provided.
Its branching ratio relative to the $J/\psi$ mode
given by the PDG is $\mathcal{B}(\Lambda_b\rightarrow \Lambda\psi(2S))/\mathcal{B}(\Lambda_b\rightarrow \Lambda J/\psi)=0.508\pm 0.023$~\cite{pdg2020}, which was deduced from the measurements by the ATLAS~\cite{plb75163} and LHCb~\cite{jhep032019126} Collaborations.
This result was lower than similar measurements in the $B$ systems,
such as $\mathcal{B}(B^0\rightarrow \psi(2S)K^0)/\mathcal{B}(B^0\rightarrow J/\psi K^0)=0.82\pm0.18$~\cite{pdg2020}
and $\mathcal{B}(B^+\rightarrow \psi(2S)K^+)/\mathcal{B}(B^+\rightarrow J/\psi K^+)=0.611\pm 0.019$~\cite{pdg2020}.
Since the uncertainty due to hadronic effects cancels to a large extent,
a comparison of the beauty meson and baryon branching ratios can be used to test the factorization
of amplitudes and provide useful information on the production of charmonia in $b$-hadron decays.

These observations motivate us to investigate the dynamics of the charmonium modes in the baryon sector.
Particularly, the decays $\Lambda_b \rightarrow \Lambda J/\psi,\Lambda \psi(2S)$ are quite appealing from a theoretical point of view in that
they proceed solely via $W$-emission diagrams, and there is no contribution due to $W$-exchange diagrams~\cite{Leibovich:2003tw}.
Meanwhile, similar  to the mesonic analog,
a significant impact of nonfactorizable contributions is expected,
which provides valuable additional information to improve our understanding of the nonfactorizable mechanism.
There have been a number of theoretical calculations for the decays under study in the literature
~\cite{Mohanta:1998iu,prd531457,prd562799,prd58014016,prd575632,mpla13181,prd80094016,ijmpa271250016,
prd95113002,prd65074030,prd99054020,prd92114013,prd96013003,prd88114018,prd92114008,plb614165,plb751127,prd98074011}.
Very recently, the angular distributions for the decays $\Lambda_b\rightarrow\Lambda^*J/\psi$,
where the $\Lambda^*$ are $\Lambda$-type excited states,
have been derived by using the helicity amplitude technique~\cite{Xing:2022uqu}.
They calculated the partial decay width, polarization, and forward-backward asymmetry.

From a theoretical point of view,
one difficult thing is to evaluate the hadronic matrix element of local operators between the initial and final states,
which require nonperturbative hadronic inputs.
The perturbative QCD factorization (PQCD) approach can serve as a useful tool for investigating the heavy baryon decays.
The basic ingredient is that the decay amplitude
is factorized into the convolution of the hard kernel, the
jet functions with the universal nonperturbative wave functions.
The jet functions organize double logarithms appearing in the hard kernel,
whose resummation gives the Sudakov factor and guarantees the removal of the end point singularities.
The PQCD approach has been developed and successfully applied to deal with various decays of a $\Lambda_b$ baryon
\cite{prd59094014,prd61114002,cjp39328,prd74034026,prd65074030,prd80034011,220204804,220209181}.
We recently have  analyzed  the  nonleptonic
decays of $\Lambda_b\rightarrow \Lambda_c \pi,\Lambda_c K$ by using the PQCD approach and  obtained satisfactory results.
This work focuses on the study of $\Lambda_b \rightarrow \Lambda J/\psi,\Lambda \psi(2S)$ decays.
The former have been studied in a previous work~\cite{prd65074030}
compared to which the analysis and scope of this work is improved in several aspects.

The $\Lambda_b$ baryon light-cone distribution amplitudes (LCDAs) are included up to the twist-4 level
according to the general Lorentz structures in Refs.~\cite{Ali:2012zza,plb665197,jhep112013191,jhep022016179}.
The LCDAs of a $\Lambda$ baryon are taken from QCD sum rules~\cite{zpc42569} at leading-twist accuracy.
For the  LCDAs of the charmonium states, we adopt
the harmonic oscillator models  proposed in our previous work~\cite{prd90114030,epjc75293},
which are successful in describing various hadronic charmonium $B$ and $B_c$ decays
\cite{epjc76564,epjc77199,epjc77610,prd97033006,prd98113003,prd99093007,epjc79792,cpc44073102,prd101016015}.
Thus,we are motivated to check for the validity of the same scenario in the baryon sector.
It is worth emphasizing that here we distinguish the LCDAs of the charmonia for the longitudinal and transverse polarizations since they exhibit different asymptotic behaviors.
In particular, the twist-3 ones contribute to the decay amplitude through the nonfactorization diagrams and
play an important role in the concerned color-suppressed decays. 
For the Sudakov factor of the charmonium states, we employ the recent updated results from Refs.~\cite{Liu:2018kuo,Liu:2020upy},
which were derived at the next-to-leading-logarithmic accuracy by including the effect from charm quark mass.
Finally, similar to the cases of hadronic charmonium $B$ meson decays,
we also consider the vertex corrections to the factorizable amplitudes at the current known next-to-leading-order level,
whose effects can be combined in the Wilson coefficients as usual~\cite{Beneke:1999br,Beneke:2000ry,Beneke:2003zv}.
In addition, the corresponding $\psi(2S)$ channel is also investigated,
which is helpful to test the factorization by its relative branching ratio as mentioned before.
Besides the decay branching ratios, many asymmetries derived from helicity amplitudes are also predicted and compared with currently available theoretical predictions and experiments.

We present our work as follows:
In Sec.~\ref{sec:framework}, we give a brief description of the theoretical framework underlying the formulation of the PQCD,
such as kinetic conventions, hadronic LCDAs,  and the effective Hamiltonian.
Thereafter, the numerical results for the transition form factors, invariant and helicity amplitudes,
decay branching ratios, up-down asymmetries, and other pertinent decay asymmetry parameters are presented in Sec.~\ref{sec:results}.
We end with a brief summary in Sec.~\ref{sec:sum}.
Appendixes~\ref{sec:LCDAs} and~\ref{sec:app} are prepared to give some details of the $\Lambda_b$ LCDAs and factorization formulas, respectively.

\section{ Theoretical framework}\label{sec:framework}
\begin{figure}[!htbh]
\begin{center}
\vspace{0.01cm} \centerline{\epsfxsize=15cm \epsffile{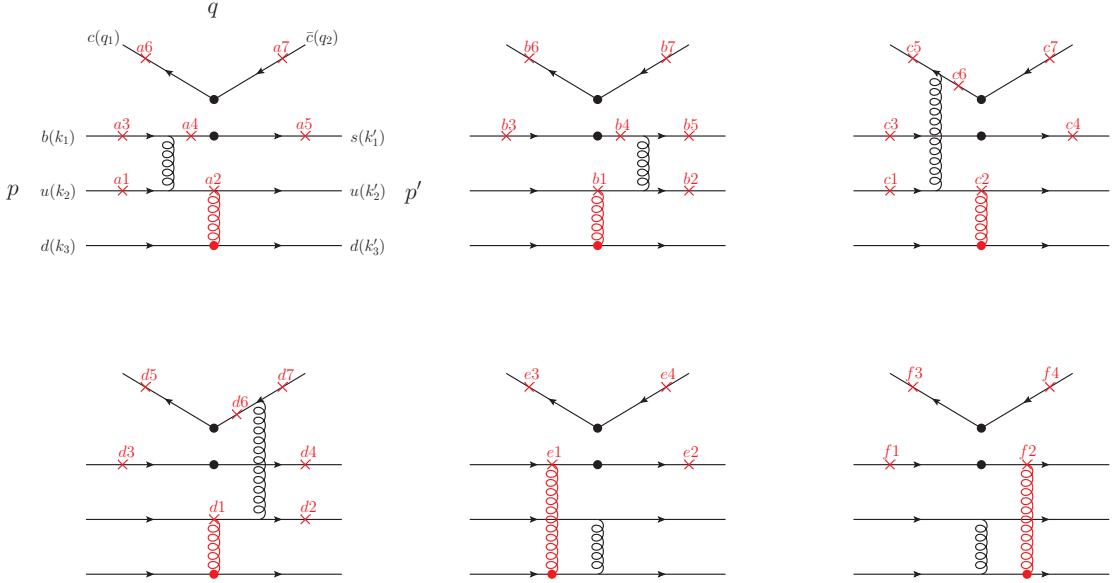}}
\setlength{\abovecaptionskip}{1.0cm}
\caption{ Feynman diagrams for the $\Lambda_b\rightarrow \Lambda J/\psi$ decay at the leading order,
where the solid black dot represents the vertex of the effective weak interaction.
The crosses on the quark lines indicated by $ij$ with $i=a-f$ and $j=1-7$ denote the possible ways in which the quark is connected to the spectator $d$ quark via a hard gluon.}
\label{fig:Feynman}
\end{center}
\end{figure}
As already mentioned, the concerned decays only receive the contributions from the $W$-emission diagrams,
in which the two spectator quarks  are shared by the parent and daughter baryons $\Lambda_b$ and $\Lambda$, respectively.
In the PQCD framework, the perturbative calculations start at the order of  $\mathcal{O}$$(\alpha_s^2)$.
The related Feynman diagrams at the leading order level are shown in Fig.~\ref{fig:Feynman}.
Following the same convention in Ref.~\cite{220209181}, each diagram is denoted by $T_{ij}$ with subscripts
$i=a-f$ and $j=1-7$ representing the possible ways of exchanging two hard gluons.
The factorizable diagrams correspond to $a1-a5$, $b1-b5$, $e1,e2$, and $f1,f2$,
while the remaining ones are all classified as nonfactorizable ones.
The triple-gluon vertex  diagrams do not contribute since the corresponding color rearrangement factors are zero in the present case.

It is convenient to work in the rest frame of the parent baryon $\Lambda_b$ with the daughter
baryon $\Lambda$ moving in the positive direction on the light cone such that
$p=\frac{M}{\sqrt{2}}(1,1,\textbf{0}_{T})$ and $p'=\frac{M}{\sqrt{2}}(f^+,f^-,\textbf{0}_{T})$ with $M$ being the $\Lambda_b$ baryon mass.
Then, the momentum ($q$) and the longitudinal and transverse polarization vectors
($\epsilon_{L,T}$) of the charmonium can be determined by
the momentum conservation and the normalization and orthogonality conditions as
\begin{eqnarray}\label{eq:pq}
q=\frac{M}{\sqrt{2}}\left(1-f^+,1-f^-,\textbf{0}_{T}\right), \quad \epsilon_L=\frac{1}{\sqrt{2}r}\left(f^+-1,1-f^-,\textbf{0}_{T}\right),
\quad \epsilon_T=\left(0,0,\textbf{1}_{T}\right),
\end{eqnarray}
where the factors
\begin{eqnarray}
f^\pm=\frac{1}{2}\left(1-r^2+r_{\Lambda}^2 \pm \sqrt{(1-r^2+r_{\Lambda }^2)^2-4r_{\Lambda}^2}\right),
\end{eqnarray}
with the mass ratio $r_{(\Lambda)}=m_{(\Lambda)}/M$, and $m_{(\Lambda)}$ is the mass of the charmonium ($\Lambda$ baryon).
The momenta of eight quarks as shown in Fig.~\ref{fig:Feynman} are parametrized as
\begin{eqnarray}
k_1&=&\left(\frac{M}{\sqrt{2}},\frac{M}{\sqrt{2}}x_1,\textbf{k}_{1T}\right),\quad
k_2=\left(0,\frac{M}{\sqrt{2}}x_2,\textbf{k}_{2T}\right),\quad
k_3=\left(0,\frac{M}{\sqrt{2}}x_3,\textbf{k}_{3T}\right),\nonumber\\
k_1'&=&\left(\frac{M}{\sqrt{2}}f^+x_1',0,\textbf{k}'_{1T}\right),\quad
k_2'=\left(\frac{M}{\sqrt{2}}f^+x_2',0,\textbf{k}'_{2T}\right),\quad
k_3'=\left(\frac{M}{\sqrt{2}}f^+x_3',0,\textbf{k}'_{3T}\right),\nonumber\\
q_1&=&\left(\frac{M}{\sqrt{2}}y(1-f^+),\frac{M}{\sqrt{2}}y(1-f^-),\textbf{q}_{T}\right),\nonumber\\
q_2&=&\left(\frac{M}{\sqrt{2}}(1-y)(1-f^+),\frac{M}{\sqrt{2}}(1-y)(1-f^-),-\textbf{q}_{T}\right),
\end{eqnarray}
where $x_{1,2,3}$, $x'_{1,2,3}$, and $y$ are the parton longitudinal momentum fractions and $\textbf{k}_{1T,2T,3T}$,
$\textbf{k}'_{1T,2T,3T}$, and $\textbf{q}_T$ are the corresponding transverse momenta.
They satisfy  the momentum conservation condition: 
\begin{eqnarray}
\sum_{l=1}^3x_l=1,\quad \sum_{l=1}^3\textbf{k}_{lT}=0.
\end{eqnarray}
A similar argument holds for the primed quantities.
Here, only the heavy $b$ and $c$ quark masses are kept, while other light quarks are treated as massless. This means only one of the dominant components of $k'_i$ and $k_{2,3}$ is kept so that $k^2=m^2\approx 0$ in the massless limit.
Since $k'_i$ are aligned with the $\Lambda$ baryon in the dominant plus direction,
their small minus components have been neglected.
The minus components of $k_2$ and $k_3$ for the soft light quarks on the $\Lambda_b$ baryon side  are selected by their inner
products with $k'_i$, which appear in the hard-kernel calculations for the concerned processes.
As the fast recoiled $\Lambda$ baryon moves approximately in the plus direction, $f^-\sim \mathcal{O}(m^2_\Lambda/M^2) $ is a small quantity, and the sum of $k'_i$  equal to $p'$ holds approximately.

In the course of the PQCD calculations, the necessary inputs contain the hadronic LCDAs of the initial and final states,
which can be constructed via the nonlocal matrix elements.
We next specify the relevant LCDAs for the present study.
After the complete classification of the three-quark LCDAs
of the $\Lambda_b$ baryon in the heavy-quark limit was constructed~\cite{plb665197},
the investigation of the $\Lambda_b$ baryon wave function has made great progress in the last decade~\cite{plb665197,jhep112013191,epjc732302,plb738334,jhep022016179,Ali:2012zza}.
Its explicit form up to twist-4 in the momentum space can be written as~\cite{220204804}
\begin{eqnarray}
(\Psi_{\Lambda_b})_{\alpha\beta\gamma}(x_i,\mu)=\frac{1}{8\sqrt{2}N_c}
\{f_{\Lambda_b}^{(1)}(\mu)[M_1(x_2,x_3)\gamma_5C^T]_{\gamma\beta}
+f_{\Lambda_b}^{(2)}(\mu)[M_2(x_2,x_3)\gamma_5C^T]_{\gamma\beta}\}[\Lambda_b(p)]_\alpha,
\end{eqnarray} 
where $\alpha,\beta,\gamma$ are the spinor indices.
$\Lambda_b(p)$ is the  heavy baryon spinor with the quantum number $I(J^P)=0(\frac{1}{2}^+)$.
$N_c$ is the number of colors. 
$C^T$ denotes the charge conjugation matrix under transpose transform.
The normalization constants $f_{\Lambda_b}^{(1)}\approx f_{\Lambda_b}^{(2)}\equiv f_{\Lambda_b}=0.030\pm0.005$ GeV$^3$~\cite{Groote:1997yr}.
The chiral-even (-odd) projector $M_{1(2)}$ reads
\begin{eqnarray}
M_1(x_2,x_3)&=&\frac{\slashed{v}\slashed{n}}{4}\Psi_3^{+-}(x_2,x_3)
+\frac{\slashed{n}\slashed{v}}{4}\Psi_3^{-+}(x_2,x_3), \nonumber\\
M_2(x_2,x_3)&=&\frac{\slashed{n}}{\sqrt{2}}\Psi_2(x_2,x_3)
+\frac{\slashed{v}}{\sqrt{2}}\Psi_4(x_2,x_3),
\end{eqnarray}
where two light-cone vectors $n=(1,0,\textbf{0}_T)$ and $v=(0,1,\textbf{0}_T)$ satisfy $n\cdot v=1$.
Here, $n$ is parallel to the four-momentum $p'$ of the $\Lambda$ baryon in the massless limit.
Several asymptotic models for the  various twist LCDAs
have been proposed in Refs.~\cite{plb665197,jhep112013191,Ali:2012zza}  and summarized in Refs.~\cite{220204804,220506095},
which are also collected in Appendix~\ref{sec:LCDAs} to make the paper self-contained.

The leading-twist $\Lambda$ baryon LCDAs have been derived using QCD sum rules~\cite{zpc42569,Farrar:1988vz}
and their higher-power corrections up to twist-6 were systematically presented in Refs.~\cite{Liu:2014uha,Liu:2008yg}.
They involve 24 LCDAs with definite twists as well as ten independent nonperturbative parameters,
which are needed to describe the local three-quark operator matrix elements.
In order to reduce the nonperturbative parameters,
we would like to adopt the LCDAs of the $\Lambda$  baryon to the leading-twist accuracy in the present work.
As will be shown in the next section, this scheme could
yield satisfactory results with fewer parameters.
The lattice QCD calculations of the leading-twist LCDAs of the full baryon octet have been performed;
the reader is referred to  Refs.~\cite{jhep020702016,prd89094511,epja55116} for details.
In terms of the notation in Ref.~\cite{prd65074030}, the nonlocal matrix element associated with the $\Lambda$ baryon is decomposed  into
\begin{eqnarray}\label{eq:wave}
(\Psi_{\Lambda})_{\alpha\beta\gamma}(k_i',\mu)
&=&\frac{1}{8\sqrt{2}N_c}
\{(\slashed{p}'C)_{\beta\gamma}[\gamma_5\Lambda({p}')]_\alpha\Phi^V(k_i',\mu)
+(\slashed{p}'\gamma_5C)_{\beta\gamma}[\Lambda({p}')]_\alpha\Phi^A(k_i',\mu) \nonumber\\
&&+(i\sigma_{\mu\nu}{p'}^\nu C)_{\beta\gamma}[\gamma^\mu\gamma_5\Lambda({p}')]_\alpha\Phi^T(k_i',\mu)\},
\end{eqnarray}
with $\sigma_{\mu\nu}=i[\gamma_\mu,\gamma_\nu]/2$.
The explicit forms of $\Phi^{V,A,T}$ at the scale $\mu=1$ GeV have been studied using QCD sum rules~\cite{zpc42569,Farrar:1988vz}.
In this work, we adopt the Chernyak-Ogloblin-Zhitnitsky (COZ) model proposed in Ref.~\cite{zpc42569}
\begin{eqnarray}\label{eq:vat}
\Phi^V(x_1,x_2,x_3)&=&42f_{\Lambda}\phi_{asy}[0.18(x_2^2-x_3^2)-0.1(x_2-x_3)],\nonumber\\
\Phi^A(x_1,x_2,x_3)&=&-42f_{\Lambda}\phi_{asy}[0.26(x_3^2+x_2^2)+0.34x_1^2-0.56x_2x_3-0.24x_1(x_2+x_3)],\nonumber\\
\Phi^T(x_1,x_2,x_3)&=&42f_{\Lambda}^T\phi_{asy}[1.2(x_3^2-x_2^2)+1.4(x_2-x_3)],
\end{eqnarray}
where $\phi_{asy}(x_1,x_2,x_3)=120x_1x_2x_3$ denotes the asymptotic form in the limit of $\mu\rightarrow \infty$.
The two normalization constants are fixed to be
$f_{\Lambda}=10f^T_{\Lambda}=6.3\times10^{-3}$ GeV$^2$~\cite{zpc42569}.
It is easy to observe that
$\Phi^{V}$ and $\Phi^{T}$ are antisymmetric under permutation of two light quarks,
but $\Phi^A$ is symmetric under the same operation.
This is  understandable because of the isospin symmetry of [ud] diquark in the $\Lambda$ baryon.
$\Phi^A$ and $\Phi^{T}$  satisfy the normalizations~\cite{zpc42569}
\begin{eqnarray}
\int_0^1\Phi^A dx_1dx_2dx_3\delta(1-x_1-x_2-x_3)=-f_{\Lambda}, \quad \int_0^1\Phi^Tx_2dx_1dx_2dx_3\delta(1-x_1-x_2-x_3)=f_{\Lambda}^T,
\end{eqnarray}
where the $\delta$ function enforces momentum conservation.

The heavy-quarkonium production mechanism is still an open question.
The nonrelativistic QCD (NRQCD) is one of the widely accepted theoretical frameworks to deal with the exclusive charmonium production processes,
for which the amplitude can be factorized into the short-distance coefficients and the NRQCD long-distance matrix elements~\cite{Bodwin:1994jh}.
In NRQCD, a charmonium state could be produced through a $c \bar c$ pair in a color-octet state plus the emission of a soft gluon~\cite{Beneke:1998ks}.
Although the color-octet matrix elements are suppressed by a factor of $v^4$ (the relative velocity between heavy quarks) in comparison to the color-singlet matrix elements,
they are compensated  by the associated larger short-distance coefficients.
Particularly,
the color-octet contributions to $J/\psi$ production in the inclusive $B$ decay were confirmed to be significant~\cite{Beneke:1998ks}.
PQCD and NRQCD are very different approaches, which employ different expansion parameters.
The DAs for the former are defined on the light cone and expanded in twist. The matrix elements for the latter are defined
in the nonrelativistic limit and expanded in $v$~\cite{Jia:2008ep}.
Therefore, the NRQCD matrix elements should not be employed in the PQCD approach.
If the color-octet contributions are included in PQCD, the soft gluon from a DA is a physical parton and must attach to a hard kernel.
They are the so-called three-parton DA contributions~\cite{Chen:2011pn,Chen:2011gv},
which are of higher twist and suppressed  by a power of $1/m_b$ with $m_b$  being the $b$ quark mass.
As stated in~\cite{Chen:2011pn}, 
a three-parton contribution is about the order of magnitude of the higher Gegenbauer
terms in the two-parton twist-3 DA, which is expected to be small.
This smallness was also consistent with the observation made in the light-cone QCD sum rules~\cite{Belyaev:1994zk}.
Therefore, we only consider the contribution from the two-parton charmonium DAs within the accuracy of the current work,  
and the color-octet contribution through the three-parton DAs will be neglected due to its smallness.

The longitudinally and transversely polarized two-parton LCDAs up to twist-3 for charmonia are decomposed into~\cite{prd71114008}
\begin{eqnarray}
\Psi_{L}&=&\frac{1}{\sqrt{2N_c}}(m\slashed{\epsilon}_L\psi^L+\slashed{\epsilon}_L\slashed{q}\psi^t),\nonumber\\
\Psi_{T}&=&\frac{1}{\sqrt{2N_c}}(m\slashed{\epsilon}_T\psi^V+\slashed{\epsilon}_T\slashed{q}\psi^T),
\end{eqnarray}
where  the  expressions of various twists $\psi^{L,T,V,t}$ have been derived~\cite{prd90114030,epjc75293}
\begin{eqnarray}
\psi^{L,T}(x,b)&=&\frac{f_\psi}{2\sqrt{2N_c}}N^{L,T}x\bar{x}\mathcal{T}(x,b)
\exp\{-\frac{m_c}{\omega}x\bar{x}[(\frac{x-\bar{x}}{2x\bar{x}})^2+\omega^2b^2]\} ,\nonumber\\
\psi^t(x,b)&=&\frac{f_\psi}{2\sqrt{2N_c}}N^t(x-\bar{x})^2\mathcal{T}(x,b)
\exp\{-\frac{m_c}{\omega}x\bar{x}[(\frac{x-\bar{x}}{2x\bar{x}})^2+\omega^2b^2]\},\nonumber\\
\psi^V(x,b)&=&\frac{f_\psi}{2\sqrt{2N_c}}N^V[1+(x-\bar{x})^2]\mathcal{T}(x,b)
\exp\{-\frac{m_c}{\omega}x\bar{x}[(\frac{x-\bar{x}}{2x\bar{x}})^2+\omega^2b^2]\},
\end{eqnarray}
with $m_c$ being the charm quark mass and $\bar x=1-x$.
We take the shape parameters $\omega=0.6$ GeV for $J/\psi$~\cite{prd90114030} and  $\omega=0.2$ GeV for $\psi(2S)$~\cite{epjc75293}.
The normalization constants $N^{L,T,V,t}$ are related to the decay constants $f_\psi$
via the normalization
\begin{eqnarray}
\int_0^1\psi^{L,T,V,t}(x,0)dx=\frac{f_\psi}{2\sqrt{2N_c}}.
\end{eqnarray}
The function $\mathcal{T}(x,b)$ reads~\cite{prd90114030,epjc75293}
\begin{align}
\begin{split}
\mathcal{T}(x,b)= \left \{
\begin{array}{ll}
1, &  \text{for} \quad J/\psi, \\
1-4b^2m_c\omega x\bar{x}+\frac{m_c(x-\bar{x})^2}{\omega x\bar{x}}, &  \text{for} \quad \psi(2S).
\end{array}
\right.
\end{split}
\end{align}

Based on the operator product expansion, the effective weak-interaction Hamiltonian for the $b\rightarrow s c\bar c$ transition reads~\cite{Buchalla:1995vs}
\begin{eqnarray}
\mathcal{H}_{eff}&=&\frac{G_F}{\sqrt{2}} \{V_{cb}V^*_{cs}[C_1(\mu)O_1(\mu)+C_2(\mu)O_2(\mu)]-\sum_{k=3}^{10}V_{tb}V^*_{ts}C_k(\mu)O_k(\mu)\}+\mathrm{H.c.},
\end{eqnarray}
where $G_F$ is the Fermi constant, $V_{ij}$ represent the Cabibbo-Kobayashi-Maskawa (CKM) matrix elements,
$C_i(\mu)$ correspond to the Wilson coefficients evaluated at the renormalization scale $\mu$,
and  $O_i$ are the four-quark operators, defined by
\begin{eqnarray}
O_1&=& \bar{c}_\alpha \gamma_\mu(1-\gamma_5) b_\beta  \otimes \bar{s}_\beta  \gamma^\mu(1-\gamma_5) c_\alpha, \nonumber\\
O_2&=& \bar{c}_\alpha \gamma_\mu(1-\gamma_5) b_\alpha \otimes \bar{s}_\beta  \gamma^\mu(1-\gamma_5) c_\beta,  \nonumber\\
O_3&=& \bar{s}_\beta  \gamma_\mu(1-\gamma_5) b_\beta  \otimes \sum_{q'} \bar{q}'_\alpha \gamma^\mu(1-\gamma_5) q'_\alpha, \nonumber\\
O_4&=& \bar{s}_\beta  \gamma_\mu(1-\gamma_5) b_\alpha \otimes \sum_{q'} \bar{q}'_\alpha \gamma^\mu(1-\gamma_5) q'_\beta,  \nonumber\\
O_5&=& \bar{s}_\beta  \gamma_\mu(1-\gamma_5) b_\beta  \otimes \sum_{q'} \bar{q}'_\alpha \gamma^\mu(1+\gamma_5) q'_\alpha, \nonumber\\
O_6&=& \bar{s}_\beta  \gamma_\mu(1-\gamma_5) b_\alpha \otimes \sum_{q'} \bar{q}'_\alpha \gamma^\mu(1+\gamma_5) q'_\beta,  \nonumber\\
O_7&=&   \frac{3}{2}\bar{s}_\beta \gamma_\mu(1-\gamma_5) b_\beta  \otimes \sum_{q'} e_{q'}\bar{q}'_\alpha \gamma^\mu(1+\gamma_5) q'_\alpha, \nonumber\\
O_8&=&   \frac{3}{2}\bar{s}_\beta \gamma_\mu(1-\gamma_5) b_\alpha \otimes \sum_{q'} e_{q'}\bar{q}'_\alpha \gamma^\mu(1+\gamma_5) q'_\beta,  \nonumber\\
O_9&=&   \frac{3}{2}\bar{s}_\beta \gamma_\mu(1-\gamma_5) b_\beta  \otimes \sum_{q'} e_{q'}\bar{q}'_\alpha \gamma^\mu(1-\gamma_5) q'_\alpha, \nonumber\\
O_{10}&=&\frac{3}{2}\bar{s}_\beta \gamma_\mu(1-\gamma_5) b_\alpha \otimes \sum_{q'} e_{q'}\bar{q}'_\alpha \gamma^\mu(1-\gamma_5) q'_\beta.
\end{eqnarray}
The sum over $q'$ runs over the quark fields that are active at the scale $\mu=\mathcal{O}(m_b)$.
It is evident that the direct $CP$ violations for the decay at hand are very small
owing to an almost null weak phase from the CKM matrix element $V_{ts}$.
The helicity amplitudes $H_{\lambda_\Lambda\lambda_\psi}$ are then given by sandwiching $\mathcal{H}_{eff}$ with the initial and final states,
\begin{eqnarray}
H_{\lambda_\Lambda\lambda_\psi}=\langle \Lambda(\lambda_\Lambda)\psi(\lambda_\psi)|\mathcal{H}_{eff}|\Lambda_b(\lambda_{\Lambda_b})\rangle,
\end{eqnarray}
where $\lambda_{\Lambda_b}$, $\lambda_{\Lambda}$, and $\lambda_{\psi}$ are the corresponding helicities.
The helicity labels can take the values $\lambda_{\Lambda}=\pm 1/2$ and $\lambda_{\psi}=0,\pm 1$.
Angular momentum conservation in the $\Lambda_b$ decay requires $|\lambda_{\Lambda}-\lambda_{\psi}|=1/2$.
Thereby, the decay can be described by four independent complex helicity amplitudes,
namely, $H_{-\frac{1}{2}-1}$, $H_{\frac{1}{2}1}$, $H_{\frac{1}{2}0}$, and $H_{-\frac{1}{2}0}$, in the helicity-based definition.
The former two terms correspond to the transverse polarizations for the charmonium and the last two terms to the longitudinal ones.
As we will see later, the helicity amplitudes are particularly convenient for expressing various observable quantities in the decays.

In general, it is convenient to analyze the decay in terms of the invariant amplitudes, 
which can be expanded with the Dirac spinors and polarization vector as
\begin{eqnarray}\label{eq:kq}
\mathcal{M}^L&=&\bar {u}_\Lambda (p')\epsilon^{\mu*}_{L}[A_1^L\gamma_{\mu}\gamma_5+A_2^L\frac{p'_{\mu}}{M}\gamma_5+B_1^L\gamma_\mu+B_2^L\frac{p'_{\mu}}{M}]u_{\Lambda_b}(p),
\nonumber\\
\mathcal{M}^T&=&\bar {u}_\Lambda (p')\epsilon^{\mu*}_{T}[A_1^T\gamma_{\mu}\gamma_5+B_1^T\gamma_\mu]u_{\Lambda_b}(p),
\end{eqnarray}
where $A^{L,T}_{1,2}$ and $B^{L,T}_{1,2}$ are the so-called invariant amplitudes
with $L$ and $T$ in the superscripts denoting the longitudinal and transverse components, respectively.
We emphasize that the two structures $A_1^L\gamma_{\mu}\gamma_5$ and $A_2^L\frac{p'_{\mu}}{M}\gamma_5$
 appearing in the above equations are not independent.
The same statement is also true for the $B_1^L\gamma_\mu$ and $B_2^L\frac{p'_{\mu}}{M}$ terms.
It is easy to prove it by expressing the longitudinal polarization vector in terms of momenta $p$ and $p'$,
and using  the equations of motion for the two spinors.
In fact, there are also four independent invariant amplitudes, as there should be.
The explicit relations between the helicity amplitudes $H_{\lambda_\Lambda\lambda_\psi}$
and the invariant amplitudes $A,B$ are~\cite{zpc55659,prd562799}
\begin{eqnarray}\label{eq:helicity}
H_{\frac{1}{2}1}&=&-(\sqrt{Q_+}A_1^T+\sqrt{Q_-}B_1^T),\nonumber\\
H_{-\frac{1}{2}-1}&=&\sqrt{Q_+}A_1^T-\sqrt{Q_-}B_1^T,\nonumber\\
H_{\frac{1}{2}0}&=& \frac{1}{\sqrt{2}m}[\sqrt{Q_+}(M-m_\Lambda)A_1^L-\sqrt{Q_-}P_cA_2^L+\sqrt{Q_-}(M+m_\Lambda)B_1^L+\sqrt{Q_+}P_cB_2^L],\nonumber\\
H_{-\frac{1}{2}0}&=& \frac{1}{\sqrt{2}m}[-\sqrt{Q_+}(M-m_\Lambda)A_1^L+\sqrt{Q_-}P_cA_2^L+\sqrt{Q_-}(M+m_\Lambda)B_1^L+\sqrt{Q_+}P_cB_2^L],
\end{eqnarray}
with $Q_{\pm}=(M\pm m_\Lambda)^2-m^2$. $P_c$ 
denotes the modulus of the three momentum of the $\Lambda$ in the $\Lambda_b$ rest frame.

The general factorization formula for any one of the invariant amplitudes in Eq.~(\ref{eq:kq}) can be  written as
\begin{eqnarray}\label{eq:amp}
\mathcal{F}=\frac{f_{\Lambda_b}\pi^2 G_F}{18\sqrt{3}}\sum_{ij}
\int[\mathcal{D}x][\mathcal{D}b]_{T_{ij}}
\alpha_s^2(t_{T_{ij}})\Omega_{T_{ij}}(b,b',b_q)e^{-S_{T_{ij}}}[V^{LL}H^{LL}_{T_{ij}}(x,x',y)+V^{SP}H^{SP}_{T_{ij}}(x,x',y)],
\end{eqnarray}
where the summation extends over all possible diagrams as shown in Fig.~\ref{fig:Feynman}.
$V$ denotes the product of the CKM matrix elements and the Wilson coefficients,
where the superscripts $LL$ and $SP$ correspond to the contributions from the (V-A)(V-A) and (S-P)(S+P) operators, respectively.
$H_{T_{ij}}$ is the numerator of the hard amplitude depending on the spin structure of the final state.
These quantities associated with specific diagram are collected in Appendix~\ref{sec:app}.
$\Omega_{T_{ij}}$ is the Fourier transformation of the denominator of the hard amplitude from the $k_T$ space to its conjugate $b$ space.
Their explicit formulas  can be found in Ref.~\cite{220209181} and shall not be repeated here.
The integration measure of the momentum fractions are defined as
\begin{eqnarray}
[\mathcal{D}x]=[dx][dx']dy, \quad [dx]=dx_1dx_2dx_3\delta(1-x_1-x_2-x_3),\quad [dx']=dx'_1dx'_2dx'_3\delta(1-x'_1-x'_2-x'_3),
\end{eqnarray}
and the measure of the transverse extents $[\mathcal{D}b]$ are also shown in Ref.~\cite{220209181}.

The Sudakov factors in Eq.~(\ref{eq:amp}) coming from the $k_T$ resummation are given by~\cite{prd80034011}
\begin{eqnarray}
S_{T_{ij}}=\sum_{l=2,3}s(w,k^-_l)+ \sum_{l=1,2,3}s(w',k'^+_l)+
\frac{8}{3}\int^{t_{T_{ij}}}_{\kappa w}\frac{d\bar \mu}{\bar \mu}\gamma_q(\alpha_s(\bar \mu))+3\int^{t_{T_{ij}}}_{\kappa w'}\frac{d\bar \mu}{\bar \mu}\gamma_q(\alpha_s(\bar \mu)),
\end{eqnarray}
\begin{eqnarray}
S_{T_{ij}}&=&\sum_{l=2,3}s(w,k^-_l)+ \sum_{l=1,2,3}s(w',k'^+_l)+ \sum_{l=1,2}s_c(w_q,q^-_l)+\nonumber\\&&
\frac{8}{3}\int^{t_{T_{ij}}}_{\kappa w}\frac{d\bar \mu}{\bar \mu}\gamma_q(\alpha_s(\bar \mu))
+3\int^{t_{T_{ij}}}_{\kappa w'}\frac{d\bar \mu}{\bar \mu}\gamma_q(\alpha_s(\bar \mu))
+2\int^{t_{T_{ij}}}_{\kappa w_q}\frac{d\bar \mu}{\bar \mu}\gamma_q(\alpha_s(\bar \mu)),
\end{eqnarray}
for the factorizable and nonfactorizable diagrams, respectively.
The explicit expressions of the function $s_{(c)}$ can be found in Refs.~\cite{Botts:1989kf,Liu:2018kuo}.
Another threshold Sudakov factor $S_t(x)$ collecting the double logarithms $\alpha_s \ln^2(x)$ to all orders
is set to 1, similar to the case of the $\Lambda_b\rightarrow p$ decays~\cite{220204804}.
The phenomenological factor $\kappa=1.14$ is adopted according to Ref.~\cite{Kundu:1998gv}.

The hard scale $t$ for each diagram is chosen as the maximal virtuality of internal particles
including the factorization scales in a hard amplitude:
\begin{eqnarray}
t_{T_{ij}}=\max(\sqrt{|t_A|},\sqrt{|t_B|},\sqrt{|t_C|},\sqrt{|t_D|},w,w',w_q),
\end{eqnarray}
where the expressions of $t_{A,B,C,D}$ are listed in Appendix~\ref{sec:app}.
The factorization scales $w$, $w'$, and $w_q$ are defined by
 \begin{eqnarray}
w^{(')}=\min(\frac{1}{b^{(')}_1},\frac{1}{b^{(')}_2},\frac{1}{b^{(')}_3}),\quad w_q=\frac{1}{b_q},
\end{eqnarray}
 with the variables
 \begin{eqnarray}
b^{(')}_1=|b^{(')}_2-b^{(')}_3|,
\end{eqnarray}
and the other $b^{(')}_l$ defined by permutation.

\section{Numerical results}\label{sec:results}
We first present all the pertinent inputs for our numerical calculations.
Various masses (GeV), lifetimes (ps), and the Wolfenstein parameters for the CKM matrix
are summarized  below~\cite{pdg2020}
\begin{eqnarray}
M&=&5.6196, \quad  m_{\Lambda}=1.116,  \quad m_b=4.8, \quad m_c=1.275, \quad m_{J/\psi}=3.097,  \quad m_{\psi(2S)}=3.686,\nonumber\\
\quad \tau&=&1.464, \quad \lambda =0.22650, \quad  A=0.790,  \quad \bar{\rho}=0.141, \quad \bar{\eta}=0.357.
\end{eqnarray}
The decay constant of $J/\psi$ has been studied using the lattice QCD method~\cite{prd102054511},
while the $\psi(2S)$ one is still not available.
Here we choose
$f_{J/\psi}=0.363^{+0.089}_{-0.088}$ GeV and $f_{\psi(2S)}=0.309\pm 0.076$ GeV
obtained from the calculation of the $S$-wave quarkonium wave functions at the origin in the $\overline{MS}$ scheme
based on nonrelativistic effective field theories~\cite{jhep120652020}.
Other nonperturbative parameters appearing in the hadron LCDAs have been specified before.

\begin{table}[!htbh]
\caption{PQCD predictions for the form factors $f_1$ and $g_1$ at $q^2=0$ of the $\Lambda_b\rightarrow \Lambda$ transition
with the various  models of the $\Lambda_b$ LCDAs.}
\label{tab:form1}
\begin{tabular}[t]{lcccccc}
\hline\hline
Form factors & Exponential model & Free-parton approximation & Gegenbauer-1 & Gegenbauer-2 &QCDSR model& Gaussian-type     \\ \hline
$f_1(0)$     & $0.095$           & $0.125$                   & $0.090$      & $0.093$      &$0.122$    & $0.004$ \\
$g_1(0)$     & $0.104$           & $0.136$                   & $0.102$      & $0.104$      &$0.129$    & $0.005$ \\
\hline\hline
\end{tabular}
\end{table}

The calculations of the $\Lambda_b\rightarrow \Lambda$ transition form factors are similar to that of
$\Lambda_b\rightarrow \Lambda_c$~\cite{220209181},  the matrix of which is induced by the $V-A$
current has the general form
\begin{eqnarray}\label{eq:form}
\langle  \Lambda(p')|\bar s \gamma^\mu b|\Lambda_b(p)\rangle &=&\bar{u}_\Lambda(p')[f_1(q^2)\gamma^\mu
-\frac{f_2(q^2)}{M}i\sigma^{\mu\nu}q_\nu+\frac{f_3(q^2)}{M}q^\mu]u_{\Lambda_b}(p),\nonumber\\
\langle  \Lambda(p')|\bar s \gamma^\mu\gamma_5 b|\Lambda_b(p)\rangle &=&\bar{u}_\Lambda(p')[g_1(q^2)\gamma^\mu
-\frac{g_2(q^2)}{M}i\sigma^{\mu\nu}q_\nu+\frac{g_3(q^2)}{M}q^\mu]\gamma_5u_{\Lambda_b}(p),
\end{eqnarray}
where $q=p-p'$ denotes the transferred momentum between the initial and final baryons.
Here, we shall concentrate on $f_1$ and $g_1$,
which can be extracted from the factorizable decay amplitudes~\cite{220209181}.
We first compare the results at zero momentum
transfer ($q^2=0$) from various models for the $\Lambda_b$ baryon LCDAs in Table~\ref{tab:form1}.
The values from the first five models are found to be of similar size, but the numbers in the last column,
which correspond to the Gaussian-type model, are at least an order of magnitude lower.
This is ascribed to the Gaussian-type model yielding a severe suppression at the endpoint region $x_1\sim 1$,
where the $b$ quark carries most of the  $\Lambda_b$ baryon momentum.
On the contrary, other models have a strong peak in that region as shown in the Fig. 2 of Ref.~\cite{220204804}.
The different behaviors at the end point region cause the results derived from the Gaussian-type model to be generally smaller.
Actually, a similar case also occurred in earlier PQCD calculations~\cite{prd80034011,prd59094014} on the $\Lambda_b \rightarrow p$ transition
form factors with the Gaussian-type $\Lambda_b$ LCDAs,
which are also quite small, at the level of $10^{-3}$.
A recent reanalysis  in PQCD~\cite{220204804} found that the values increased significantly
by using other models instead of the Gaussian-type one.

We also examine the effect of different $\Lambda$ baryon LCDAs on the numerical results.
As aforementioned, there are another two alternative models for the $\Lambda$  baryon LCDAs:
one from the light-cone QCD sum rules (QCDSR)~\cite{Liu:2014uha} including the higher-power corrections up to
 twist-6 and the other from the lattice QCD~\cite{jhep020702016,epja55116} with  leading-twist accuracy.
Using the inputs in Ref.~\cite{Liu:2014uha},
we obtain the form factors with only the twist-3 contributions $f_1(0)=2.1$ and $g_1(0)=1.8$,
which were incredibly large.
The reason is that the coefficient of the linear term of Eq. (46) in Ref.~\cite{Liu:2014uha}
is 2 orders of magnitude greater than the analogous term in Eq.~(\ref{eq:vat}).
Taking into account contributions up to twist-6 LCDAs,
the corresponding values above become $-5.2$ and $-0.08$, respectively,
which seems to contradict with the heavy-quark symmetry.
As stressed in Refs.~\cite{jhep020702016,epja55116},
the shape parameters from the lattice QCD calculations are generally
several factors below the COZ estimates in Ref.~\cite{zpc42569},
resulting in much smaller form factors when utilizing the lattice QCD parameters.
The values may be enhanced by including contributions from higher twist LCDAs but which are currently unavailable in lattice QCD.
Hence, to obtain a reasonable estimate under the PQCD framework,
we will employ the exponential model~\cite{jhep112013191} for the $\Lambda_b$ baryon LCDAs
and the leading-twist $\Lambda$ baryon LCDAs with the corresponding parameters from the COZ model~\cite{zpc42569} in the subsequent analysis.

\begin{table}[!htbh]
\footnotesize
\caption{Theoretical predictions for the form factors $f_1$ and $g_1$ at $q^2=0$ of the $\Lambda_b\rightarrow \Lambda$ transition.}
\label{tab:form2}
\resizebox{\textwidth}{!}{
\begin{tabular}[t]{lccccccccccccc}
\hline\hline
Form factors & This work & \cite{prd80094016} & \cite{prd88114018,Gutsche:2013pp} & \cite{Mohanta:2000nk} & \cite{jhep022016179} & \cite{Detmold:2016pkz} & \cite{Aliev:2010uy} & \cite{Huang:1998ek} & \cite{ijmpa271250016} & \cite{prd99054020} & \cite{plb751127} & \cite{prd562799}                       \\ \hline
$f_1(0)$ &$0.095^{+0.057+0.018}_{-0.029-0.021}$       & $0.1081$           & $0.107$            & $0.061$               & $0.18\pm0.04$        & $\sim0.20$                 & $0.322\pm0.112$     & $0.446$             & $0.025$                & $0.131_{-0.018-0.008}^{+0.016+0.008}$ & $0.175\pm0.106$ &$0.062$ \\
$g_1(0)$   &$0.104^{+0.060+0.016}_{-0.033-0.020}$      & $0.1065$           & $0.104$            & $0.107$               & $\cdots$             & $\cdots$               & $0.318\pm0.110$     & $0.446$             & $0.028$                & $0.132_{-0.017-0.009}^{+0.016+0.008}$ & $\cdots$ &$0.108$ \\
\hline\hline
\end{tabular}}
\end{table}

In Table~\ref{tab:form2}, we compare our results on the form factors at $q^2=0$  to those obtained in other works,
where the first and second uncertainties arise
from the shape parameter $\omega_0=0.4\pm0.1$ in the $\Lambda_b$ baryon LCDAs for the exponential model
and hard scale $t$ varying from  $0.75t$  to $1.25t$, respectively.
It is found that the predominant uncertainties in our calculations stem from the baryon LCDAs,
which can reach $50\%$ in magnitude.
The relevant nonperturbative parameters in the LCDAs also need to be constrained to further improve the precision of theoretical predictions.
Broadly, our results agree reasonably well with
the results from the light-front quark model~\cite{prd80094016}, covariant confined quark model (CCQM)~\cite{prd88114018},
and the covariant light-front quark model~\cite{prd99054020} with the diquark approximation.
The values from the QCDSR~\cite{Aliev:2010uy} and Heavy Quark Effective Theory (HQET)~\cite{Huang:1998ek} are several times larger
but still at the same order.
However, the numbers from the nonrelativistic quark model~\cite{ijmpa271250016} are quite small, at the order of $10^{-2}$.
Despite a wide range of various predictions,
the nearly equal relation $f_1(0)\sim g_1(0)$ holds for the most approaches,
which is expected in the heavy-quark limit.

\begin{table}
\caption{The values of the invariant amplitudes from the factorizable and nonfactorizable diagrams for $\Lambda_b\rightarrow \Lambda \psi$ decays. Only central values are presented here.}
\label{tab:fac}
\begin{tabular}[t]{lcc}
\hline\hline
Amplitude                                     &Factorizable                                 &Nonfactorizable                             \\ \hline
$A^L_1(\Lambda_b\rightarrow \Lambda J/\psi)$    & $3.57\times 10^{-11}+i3.55\times 10^{-10}$  & $8.12\times 10^{-9}-i2.50\times 10^{-8}$   \\
$B^L_1(\Lambda_b\rightarrow \Lambda J/\psi)$    & $-6.06\times 10^{-11}-i4.17\times 10^{-10}$ & $-6.87\times 10^{-9}+i2.28\times 10^{-8}$  \\
$A^L_2(\Lambda_b\rightarrow \Lambda J/\psi)$    & $-1.32\times 10^{-11}-i8.66\times 10^{-11}$ & $2.22\times 10^{-10}+i7.35\times 10^{-9}$  \\
$B^L_2(\Lambda_b\rightarrow \Lambda J/\psi)$    & $1.81\times 10^{-11}+i1.30\times 10^{-10}$  & $8.08\times 10^{-9}-i8.59\times 10^{-9}$   \\
$A_1^T(\Lambda_b\rightarrow \Lambda J/\psi)$  & $5.79\times 10^{-11}+i3.29\times 10^{-10}$  & $7.41\times 10^{-9}-i2.95\times 10^{-8}$   \\
$B_1^T(\Lambda_b\rightarrow \Lambda J/\psi)$  & $-8.42\times 10^{-11}-i3.91\times 10^{-10}$ & $-6.82\times 10^{-9}+i2.90\times 10^{-8}$  \\
$A^L_1(\Lambda_b\rightarrow \Lambda \psi(2S))$  & $1.01\times 10^{-10}+i5.14\times 10^{-10}$  & $8.69\times 10^{-9}-i2.85\times 10^{-8}$   \\
$B^L_1(\Lambda_b\rightarrow \Lambda \psi(2S))$  & $-1.60\times 10^{-10}-i6.15\times 10^{-10}$ & $-8.57\times 10^{-9}+i2.68\times 10^{-8}$  \\
$A^L_2(\Lambda_b\rightarrow \Lambda \psi(2S))$  & $-3.45\times 10^{-11}-i1.43\times 10^{-10}$ & $3.04\times 10^{-9}+i1.63\times 10^{-8}$   \\
$B^L_2(\Lambda_b\rightarrow \Lambda \psi(2S))$  & $5.35\times 10^{-11}+i2.25\times 10^{-10}$  & $9.12\times 10^{-9}-i1.06\times 10^{-8}$   \\
$A_1^T(\Lambda_b\rightarrow \Lambda \psi(2S))$& $1.41\times 10^{-10}+i4.60\times 10^{-10}$  & $6.90\times 10^{-9}-i2.25\times 10^{-8}$   \\
$B_1^T(\Lambda_b\rightarrow \Lambda \psi(2S))$& $-2.01\times 10^{-10}-i5.53\times 10^{-10}$ & $-6.83\times 10^{-9}+i2.24\times 10^{-8}$  \\
\hline\hline
\end{tabular}
\end{table}

\begin{table}[!htbh]
\caption{
The  magnitude squared of normalized helicity amplitudes  for the $\Lambda_b\rightarrow \Lambda \psi$ decays,
where the numbers in parenthesises are the corresponding phases defined in the range $[-\pi,\pi]$ rad.
The sources of the theoretical errors are the same as in Table~\ref{tab:form2} but added in quadrature.
For comparison, the experimental data~\cite{plb72427,prd89092009,prd97072010} as well as the available predictions from CCQM~\cite{prd98074011,prd88114018} are also presented.
The superscript ``2" for the ATLAS results means square.}
\label{tab:helicity}
\begin{tabular}[t]{lcccc}
\hline\hline
Mode  &$|\hat{H}_{\frac{1}{2}1}|^2$ & $|\hat{H}_{-\frac{1}{2}-1}|^2$ & $|\hat{H}_{\frac{1}{2}0}|^2$ & $|\hat{H}_{-\frac{1}{2}0}|^2$ \\ \hline
$\Lambda_b\rightarrow\Lambda J/\psi$  &&&&\\
This work   & $0.044_{-0.007}^{+0.010}(1.84_{-0.09}^{+0.06})$  & $0.484_{-0.035}^{+0.041}(-1.32_{-0.01}^{+0.04})$
& $0.037_{-0.003}^{+0.014}(-1.11_{-0.36}^{+0.00})$ & $0.435_{-0.063}^{+0.041}(1.82_{-0.05}^{+0.01})$\\
CCQM~\cite{prd98074011}              & $2.34\times10^{-3}$  & $0.465$  & $3.24\times10^{-4}$   & $0.532$   \\
CCQM~\cite{prd88114018}              & $3.1\times10^{-3}$  & $0.47$  & $4.6\times10^{-4}$   & $0.53$   \\
LHCb~\cite{plb72427}  &$-0.10\pm0.04\pm0.03$ &$0.51\pm0.05\pm0.02$ &$0.01\pm0.04\pm0.03$&$0.57\pm0.06\pm0.03$\\
ATLAS~\cite{prd89092009}  &$(0.08^{+0.13}_{-0.08}\pm0.06)^2$ &$(0.79^{+0.04}_{-0.05}\pm0.02)^2$ &$(0.17^{+0.12}_{-0.17}\pm0.09)^2$&$(0.59^{+0.06}_{-0.07}\pm0.03)^2$\\
CMS~\cite{prd97072010}  &$0.05\pm0.04\pm0.04$ &$0.52\pm0.04\pm0.04$ &$-0.10\pm0.04\pm0.04$&$0.51\pm0.03\pm0.04$\\
$\Lambda_b\rightarrow\Lambda\psi(2S)$ &&&&\\
This work& $0.051_{-0.004}^{+0.003}(1.88_{-0.05}^{+0.04})$ & $0.365_{-0.020}^{+0.028}(-1.26_{-0.05}^{+0.02})$
& $0.061_{-0.011}^{+0.003}(-1.25_{-0.09}^{+0.03})$ & $0.523_{-0.030}^{+0.035}(1.83_{-0.00}^{+0.06})$\\
CCQM~\cite{prd88114018}              & $1.2\times10^{-2}$  & $0.54$  & $3.3\times10^{-3}$   & $0.45$   \\
\hline\hline
\end{tabular}
\end{table}

Next, we turn to the decay amplitudes of  the concerned decays.
The factorizable and nonfactorizable contributions to the invariant amplitudes are presented in Table~\ref{tab:fac}.
Note that the imaginary parts of the factorizable amplitudes arise due to the vertex corrections.
The decay amplitudes are clearly governed by the imaginary part of the nonfactorizable contributions.
 The factorizable contributions are smaller by 1 or 2 orders of magnitude,
 despite receiving the enhancement from the vertex corrections.
Previous PQCD investigations~\cite{prd65074030} have also observed a similar feature.
This situation differs from the case of color-suppressed decays in the $B$ meson sector,
where factorizable diagram contributions could be comparable to nonfactorizable ones
after including the vertex corrections~\cite{epjc77610,prd89094010}.

Following Ref.~\cite{prd98074011}, we introduce the moduli squared of normalized helicity amplitudes
$|\hat{H}_{\lambda_\Lambda\lambda_\psi}|^2=|H_{\lambda_\Lambda\lambda_\psi}|^2/H_N$ with
\begin{eqnarray}\label{eq:hn}
H_N=|H_{\frac{1}{2}1}|^2+|H_{-\frac{1}{2}-1}|^2+|H_{\frac{1}{2}0}|^2+|H_{-\frac{1}{2}0}|^2.
\end{eqnarray}
The calculation of the normalized squared helicity amplitudes can be done in a straightforward way by
combining Eqs.~(\ref{eq:helicity}) and ~(\ref{eq:hn}) and Table~\ref{tab:fac},
whose numerical results are exhibited in Table~\ref{tab:helicity}.
The values given in the parentheses are the corresponding phases.
The sources of the errors in the numerical estimates have the same origin as in
the discussion of the form factors in Table~\ref{tab:form2}.
Note that these quantities are less sensitive to the considered uncertainties
since the errors partially cancel in the ratios;
 thus,we have added them in quadrature.
It is observed that the considered decays receive significant contributions from
$H_{-\frac{1}{2}-1}$ and $H_{-\frac{1}{2}0}$,
which means the negative-helicity states for the $\Lambda$ baryon are preferred.
Contributions from the $\lambda_{\Lambda}=\frac{1}{2}$ helicity states are rather small,
which amounts to less than $10\%$.
This pattern  is consistent with the expectation from the heavy-quark limit and the left-handed nature of the weak interaction~\cite{jhep062020110}.
Experimentally,
a fit to the angular distribution of the cascade decay $\Lambda_b\rightarrow \Lambda(p\pi^-) J/\psi(\mu^+\mu^-)$ has been performed by the
LHCb~\cite{plb72427}, ATLAS~\cite{prd89092009}, and CMS~\cite{CMS:2016iaf,prd97072010} Collaborations,
which allows us to determine the helicity amplitudes from the fitted angular parameters.
The measured results together with the predictions from the CCQM~\cite{prd98074011,prd88114018}
are also presented in Table~\ref{tab:helicity} for comparison.
Our results are comparable with CCQM calculations and experiments.
We note that some determined helicity amplitudes squared from the LHCb and CMS experiments are  negative and therefore nonphysical.
This is because they use an old average value of $\alpha_{\Lambda}=0.642\pm0.013$ from Refs.~\cite{Astbury:1975hn,Cleland:1972fa,Dauber:1969hg,Overseth:1967zz,Cronin:1963zb}, %
which is smaller than
the recent measurements by the BES III~\cite{BESIII:2018cnd} and LHCb~\cite{jhep062020110} Collaborations
and an independent estimate of kaon photoproduction scattering data by CLAS Collaboration in Ref.~\cite{Ireland:2019uja}.

The only available experimental information about the phase of the helicity amplitudes for the $\Lambda_b\rightarrow \Lambda J/\psi$ decay comes from
the recent measurements by LHCb~\cite{jhep062020110}.
The phase of $b_+$, which corresponds to our $\text{arg}(\hat{H}_{-\frac{1}{2}-1})$, is fixed to be zero,
while the phases of the remaining three helicity amplitudes are measured relative to $\text{arg}(b_+)$.
Following the definition given in Ref.~\cite{jhep062020110},
one can estimate the three relative phases (in rad) from Table~\ref{tab:helicity} to be
\begin{eqnarray}
\text{arg}(a_+)=0.22^{+0.00}_{-0.38}(0.01^{+0.06}_{-0.05}), \quad \text{arg}(a_-)=-3.13^{+0.00}_{-0.07}(3.09^{+0.10}_{-0.01}), \quad \text{arg}(b_-)=-3.12_{-0.08}^{+0.03}(3.14^{+0.05}_{-0.02}),
\end{eqnarray}
where the corresponding numbers for the $\psi(2S)$ mode are shown in parentheses.
Because the imaginary part of the amplitude contributes significantly more than the real part as mentioned before,
these amplitude vectors are scattered toward the imaginary axis of the complex plane,
resulting in phase differences that either approach zero or $\pm\pi$.
Although our central values are away from the  most probable values from the LHCb experiment,
they still fall into the $95\%$ credibility intervals.
Since the measured phases  span  wide intervals,
we cannot draw anymore definite conclusions.
A measurement with improved precision
 on the phases would be highly desirable.

Armed with the helicity amplitudes derived above, we can now proceed
to perform the calculations of the decay branching ratios and various asymmetry parameters,
which are defined as~\cite{prd98074011,plb72427}
\begin{eqnarray}\label{eq:alp}
\mathcal{B}&=&\frac{P_c\tau_{\Lambda_b}}{8\pi M^2}H_N, \nonumber\\
\alpha_b &=&-|\hat{H}_{\frac{1}{2}1}|^2+|\hat{H}_{-\frac{1}{2}-1}|^2+|\hat{H}_{\frac{1}{2}0}|^2-|\hat{H}_{-\frac{1}{2}0}|^2, \nonumber\\
\alpha_2&=&|\hat{H}_{\frac{1}{2}1}|^2-|\hat{H}_{-\frac{1}{2}-1}|^2+|\hat{H}_{\frac{1}{2}0}|^2-|\hat{H}_{-\frac{1}{2}0}|^2, \nonumber\\
r_0 &=&|\hat{H}_{\frac{1}{2}1}|^2+|\hat{H}_{-\frac{1}{2}0}|^2, \nonumber\\
r_1 &=&|\hat{H}_{\frac{1}{2}1}|^2-|\hat{H}_{-\frac{1}{2}0}|^2,
\end{eqnarray}
where $\alpha_b$ is the parity-violating asymmetry parameter, 
$\alpha_2$ represents the longitudinal polarization of the daughter $\Lambda$ baryon,
and $r_0 (r_1)$ corresponds to the longitudinal unpolarized (polarized) parameter.
\begin{table}[!htbh]
\caption{Branching ratios ($10^{-4}$) and  asymmetry paramaters for the $\Lambda_b\rightarrow \Lambda \psi$ decays.
The sources of our theoretical errors are the same as in Table~\ref{tab:form2}.}
\label{tab:asym}
\begin{tabular}[t]{lccccc}
\hline\hline
Mode & $\mathcal{B}$ & $\alpha_b$ & $\alpha_2$  &$r_0$ &$r_1$\\ \hline
$\Lambda_b\rightarrow\Lambda J/\psi$  &&&&&\\
This work & $7.75_{-3.17-0.96}^{+4.08+1.39}$ & $0.042_{-0.069-0.001}^{+0.104+0.027}$ & $-0.84_{-0.01-0.00}^{+0.05+0.01}$ & $0.47_{-0.05-0.02}^{+0.04+0.00}$ & $-0.40_{-0.04-0.00}^{+0.07+0.01}$ \\
CCQM~\cite{prd98074011,prd96013003}  & $8.0$ & $-0.069$ & $-0.995$ & $0.533$  & $-0.532$ \\
LHCb~\cite{plb72427} &$\cdots$ & $0.05\pm0.17\pm0.17$ &$\cdots$ &$0.58\pm0.02\pm0.01$ &$-0.56\pm0.10\pm0.05$ \\
ATLAS~\cite{prd89092009} &$\cdots$ & $0.30\pm0.16\pm0.16$ &$\cdots$ &$\cdots$ &$\cdots$\\
CMS~\cite{prd97072010} &$\cdots$ & $-0.14\pm0.14\pm0.10$  & $-1.11\pm0.04\pm0.05$ &$\cdots$ &$\cdots$\\
$\Lambda_b\rightarrow\Lambda\psi(2S)$ &&&&&\\
This work& $3.63_{-1.39-0.41}^{+1.91+0.70}$ & $-0.149_{-0.048-0.040}^{+0.033+0.044}$  & $-0.78_{-0.02-0.02}^{+0.00+0.01}$ & $0.58_{-0.02-0.02}^{+0.02+0.02}$ & $-0.46_{-0.04-0.03}^{+0.02+0.03}$  \\
CCQM~\cite{prd88114018}               & $7.25$ & $0.09$&$-0.97$& 0.45 & $-0.44$    \\
\hline\hline
\end{tabular}
\end{table}
The numerical results are collected in Table~\ref{tab:asym} in comparison with CCQM calculations and
available experimental data~\cite{prd98074011,prd96013003,prd88114018}.
Some important features of the numerical results are the following:
\begin{enumerate}
\item
The obtained branching ratios for the two modes are both of order $10^{-4}$.
As aforementioned, there is no experimental data on the absolute branching ratios
due to the fact that the fragmentation fraction of the $b$ quark to $\Lambda_b$ baryon are not well determined yet.
Using the estimates of the fragmentation fraction $f(b\rightarrow\Lambda_b)= (7-17.5)\%$
from Refs.~\cite{jhep110672015,plb751127,Jiang:2018iqa},
we can convert the data in Eq.~(\ref{eq:bbb}) into the possible range
$\mathcal{B}(\Lambda_b\rightarrow\Lambda J/\psi)-(3.3-8.3)\times 10^{-4}$,
which overlaps our central value in Table~\ref{tab:asym}.
Likewise, no absolute measurement of the decay rate for the $\psi(2S)$ decay is provided,
but its relative rate with respect to the $J/\psi$ one has been measured
by the ATLAS~\cite{plb75163} and LHCb~\cite{jhep032019126} Collaborations.
The ratio in PQCD  is estimated to be
$\mathcal{R}=\frac{\mathcal{B}(\Lambda_b\rightarrow\Lambda \psi(2S))}{\mathcal{B}(\Lambda_b\rightarrow\Lambda J/\psi)}=0.47^{+0.02}_{-0.00}$,
where all uncertainties are added in quadrature.
It is apparent that this ratio is less sensitive to the variations of
hadronic parameters involved in the baryon LCDAs than the individual branching ratios,
since the parameter dependence of the PQCD predictions for the
branching ratios are largely canceled in their relative ratios.
Our prediction is consistent with the PDG average value of $\mathcal{R}=0.508\pm 0.023$~\cite{pdg2020} within uncertainties.
Although the predicted  branching ratio $\mathcal{B}(\Lambda_b\rightarrow\Lambda J/\psi)$
from CCQM agrees well with our result, its value for the $\psi(2S)$ mode one is twice larger,
leading to a larger value $\mathcal{R}=0.8\pm0.1$~\cite{prd88114018,prd92114008}.
In addition, Wei \textit{et al.}~\cite{prd80094016} used the light-front
quark model by treating the spectator quarks as a diquark system and found $\mathcal{R}=0.65$.
Mott and Roberts~\cite{ijmpa271250016}  also studied the same topic  in a nonrelativistic quark model.
They estimated the ratio even exceeding 1, which seems to be unexpected
since the decay process involving the radial excited states in the final state usually has  a lower rate.
 This demonstrates that the currently available theoretical calculations of the ratio  are generally  greater  than
our result as well as the data.
\item
It is observed from Table~\ref{tab:asym} that the predicted up-down asymmetries for the two modes seem to be distinctly different.
For the $J/\psi$ mode, we obtained the small positive central value $\alpha_b=0.042$,
while that of the $\psi(2S)$ mode is larger in size but with a minus sign. 
According to  the definition of the $\alpha_b$ in Eq.~(\ref{eq:alp}),
its value is sensitive to  the difference of  two dominant helicity amplitudes
$|\hat{H}_{-\frac{1}{2}-1}|$ and $|\hat{H}_{-\frac{1}{2}0}|$.
As can be seen from Table~\ref{tab:fac},
the transverse components are reduced going from the $J/\psi$  mode to the $\psi(2S)$ mode,
but the longitudinal ones are the opposite.
It is understandable since the dominant nonfactorizable contributions are process dependent.
A negative up-down asymmetry appears when the longitudinal polarization amplitude exceeds the transverse one.
The previous measurements of $\alpha_b$ by LHCb~\cite{plb72427}, ATLAS~\cite{prd89092009},
and CMS~\cite{prd97072010} for the $J/\psi$ channel have large errors. 
The central values given by the LHCb~\cite{plb72427} and ATLAS~\cite{prd89092009} are both positive,
while CMS~\cite{prd97072010} reports a negative one as shown in Table~\ref{tab:asym}.
Measurements that are more precise have been released recently by LHCb~\cite{jhep062020110} using data
collected with the LHCb experiment during Runs 1 and 2 of the LHC.
The resulting  most probable value of $\alpha_b$ is $-0.022$
with a $68\%$ credibility interval from  $-0.048$ to $0.005$,
which covers our calculation with error bars.

\item
The $\Lambda$ longitudinal polarizations $\alpha_2$ are significantly smaller
than those calculations from CCQM~\cite{prd98074011,prd96013003,prd88114018}.
From Table~\ref{tab:helicity}, one can see the contributions from the positive-helicity
states in CCQM are strongly suppressed at the level of $10^{-3}$.
Nonetheless, the corresponding quantities in PQCD have a order of $10^{-2}$
due to the inclusion of the nonfactorizable contributions.
The CCQM calculations are based on the factorization approximation,
in which  the hadronic matrix element for a two-body baryon decay process is factorized into
a product of a baryonic transition form factor and the meson decay constant.
The nonfactorizable effects enter into the effective Wilson coefficients
in the scenario of the effective color number,
which can be totally factorized out within the factorization framework
and canceled in the longitudinal polarizations $\alpha_2$.
In contrast, the nonfactorizable contributions in the PQCD calculations are related to the decay processes
and are also helicity dependent~\cite{prd71114008}.
In the absence of the nonfactorizable contributions,
the PQCD predictions of the $\Lambda$ longitudinal polarizations for the $J/\psi$ and $\psi(2S)$ modes will be increased to
$-0.94$ and $-0.90$, respectively,
which are closer to the CCQM results.
As a by-product, the obtained value of $\alpha_2$
allows us to estimate the forward-backward (FB) asymmetry with respect to the hadron-side polar angle~\cite{prd88114018}
$\mathcal{A}_{FB}\equiv \alpha_2\alpha_{\Lambda}=-0.613^{+0.045}_{-0.022}(-0.568^{+0.016}_{-0.033})$,
where the number in parentheses is the corresponding value for the $\psi(2S)$ mode.
In the estimates, we have used the new experimental PDG average value for the asymmetry parameter
 $\alpha_{\Lambda}= 0.732 \pm0.014$~\cite{pdg2020}.
The PQCD predictions of the FB asymmetry above can be confronted with data in the future.

\item
The predicted longitudinal unpolarized $r_0$ and polarized $r_1$ are
comparable with the CCQM calculations~\cite{prd98074011,prd96013003}
and the reported values by the LHCb Collaboration~\cite{plb72427} within errors.

\item
There is another interesting parameter $\gamma_0$ defined via~\cite{prd88114018}
\begin{eqnarray}
\gamma_0&=&|\hat{H}_{\frac{1}{2}1}|^2+|\hat{H}_{-\frac{1}{2}-1}|^2-2|\hat{H}_{\frac{1}{2}0}|^2-2|\hat{H}_{-\frac{1}{2}0}|^2,
\end{eqnarray}
which denotes the longitudinal/transverse composition of the charmonium meson.
The exact numbers in PQCD are $-0.416^{+0.089}_{-0.064}$  and  $-0.753^{+0.056}_{-0.044}$ for the $J/\psi$ and $\psi(2S)$ modes, respectively.
The former differs by $1.6\sigma$ from the CMS measurement $-0.27\pm0.08\pm0.11$~\cite{prd97072010}
but matches its previous value $-0.46\pm0.07\pm0.04$~\cite{CMS:2016iaf} within $1.0\sigma$.
A measurement with improved precision helps to better understand this possible discrepancy.
\end{enumerate}

As previously stated, many other studies have been conducted of the $\Lambda_b\rightarrow \Lambda J/\psi$ decay but have primarily focused on the decay branching ratio and the up-down asymmetry.
For the sake of comparison, we briefly summarize the currently available theoretical results in Table~\ref{tab:branching2}.
Most of the various approaches give predictions of the same order of magnitude for the decay branching ratio.
Our results coincide with those of the quark model in the large $N_c$ limit~\cite{prd95113002} and the nonrelativistic quark model~\cite{ijmpa271250016}
but generally higher than other predictions.
On the other hand, our prediction of the up-down asymmetry for the $J/\psi$ mode
is small in size with positive central value.
A variety of quark-model-based analyses and predictions gave negative values ranging from $-0.09$ to $-0.21$,
whereas the calculation of HQET provided a sizable positive value of $0.77$~\cite{Leitner:2004dt,plb614165}.
Hence, an accurate measurement of the branching ratio and up-down asymmetry parameter will enable us to discern different model predictions.
We mention that our branching ratio and the up-down asymmetry for the $J/\psi$ mode
differ from those of the previous PQCD calculations~\cite{prd65074030}
mainly because we updated the nonperturbation hadron LCDAs as discussed at length in the previous section.

\begin{table}[!htbh]
\footnotesize
\caption{
Various predictions in the literature of the $\Lambda_b\rightarrow \Lambda J/\psi$ decay,
 where the branching ratio is in units of $10^{-4}$.}
\label{tab:branching2}
\begin{tabular}[t]{lcccccccccccc}
\hline\hline
          & \cite{Mohanta:1998iu} & \cite{prd562799} & \cite{prd58014016}& \cite{prd575632,mpla13181} & \cite{prd80094016} & \cite{ijmpa271250016}
         & \cite{prd95113002}~\footnotemark& \cite{prd65074030} & \cite{prd99054020} & \cite{prd531457} & \cite{prd92114013,plb751127} & \cite{Leitner:2004dt,plb614165}~\footnotemark    \\ \hline
$\mathcal{B}$    & $2.49$& $1.6$ & $6.04$& $2.7$ & $3.94$ & $8.2$& $5.0-7.8$
& $1.65-5.27$ & $3.33_{-0.20-0.63-0.30}^{+0.48+0.56+0.32}$& $2.1$&$3.3\pm2.0$
& $12.5/4.4/1.2$  \\
$\alpha_b$ & $-0.208$& $-0.10$ & $-0.18$& $-0.21$ & $-0.204$ & $-0.09$ & $\cdots$ & $-0.17\sim-0.14$ & $-0.21\pm0.00$ & $-0.11$& $\cdots$
& $0.777$    \\
\hline\hline
\end{tabular}
\footnotetext[1]{We quote the values in the large $N_c$ limit. }
\footnotetext[2]{The quoted values correspond to $N_c^{\text{eff}}=2, 2.5$, and 3, respectively.}
\end{table}

Finally, we discuss the long-distance (LD) effects on the decay rate for the semileptonic
$\Lambda_b\rightarrow \Lambda l^+l^-$ decays with $l=e,\mu,\tau$
in the vicinity of the charmonium resonance regions
defined by~\cite{epjc59861}
\begin{eqnarray}
\mathcal{B}_{\text{LD}}(\Lambda_b\rightarrow \Lambda  l^+l^-)
=\mathcal{B}(\Lambda_b\rightarrow \Lambda J/\psi) \times \mathcal{B}(J/\psi\rightarrow l^+l^-)+
 \mathcal{B}(\Lambda_b\rightarrow \Lambda \psi(2S)) \times \mathcal{B}(\psi(2S)\rightarrow l^+l^-),
\end{eqnarray}
where other higher radial excitations are not included here since their dileptonic decay rates suffer strong suppression.
Utilizing the experimental PDG average values  $\mathcal{B}(\psi\rightarrow l^+l^-)$~\cite{pdg2020},
\begin{eqnarray}
\mathcal{B}(J/\psi\rightarrow \mu^+\mu^-)&=&
(5.961\pm0.033)\%,  \nonumber\\ 
\mathcal{B}(\psi(2S)\rightarrow l^+l^-)&=&
\left\{
\begin{aligned}
&(8.0\pm0.6)\times 10^{-3},   &l=\mu,\\
&(3.1\pm0.4)\times 10^{-3},   &l=\tau,\\
\end{aligned}\right.
\end{eqnarray}
we have
\begin{eqnarray}
\mathcal{B}_{\text{LD}}(\Lambda_b\rightarrow \Lambda l^+l^-)=
\left\{
\begin{aligned}
&(4.91_{-2.12}^{+2.78})\times 10^{-5},   &l=\mu \\
&(1.12_{-0.54}^{+0.86})\times 10^{-6},   &l=\tau,\\
\end{aligned}\right.
\end{eqnarray}
where we present the results for the $\mu$ and $\tau$ channels only
since the values for the $e$ and $\mu$ channels are essentially the same.
It is clear that the LD effects in the tau lepton mode are rather smaller than the counterpart  for the $\mu$ channel
because the secondary process $J/\psi\rightarrow \tau^+\tau^-$ is kinematically forbidden.
Our result for the $\mu$ channel is comparable with those in the light-cone sum rules~\cite{epjc59861},
QCD sum rules, the pole model~\cite{prd63114024}, and several supersymmetric scenarios evaluated in Refs.~\cite{prd78114032,ijmpa271250016}.
However, the observation is different for the semitauonic process,
for which the obtained results from various approaches span a wide range $(0.59-11)\times 10^{-6}$~\cite{epjc59861,prd63114024,prd78114032,ijmpa271250016}.
Future experimental data are expected to verify and distinguish the various results.

\section{ conclusion}\label{sec:sum}

Thanks to the continuous efforts of the LHC experiment,
a large data sample of $b$ hadrons are produced,
offering a unique opportunity to study $\Lambda_b$ weak decay systematically.
In this work, we have carefully explored the color-suppressed baryonic decays
$\Lambda_b\rightarrow \Lambda J/\psi,\Lambda \psi(2S)$ in the framework of PQCD,
for which the decay process is factorized into the calculable hard kernel and universal hadron distribution amplitudes.
In light of  these improved universal nonperturbative objects,
we have calculated  some observables related to the decays under consideration
including both the factorizable and nonfactorizable contributions.
Our main results are summarized as follows:
\begin{itemize}
\item[(I)]
Five phenomenological models  for the LCDAs of the $\Lambda_b$ baryon have been employed for comparison.
It has been found that the form factors at maximum recoil evaluated by employing the QCDSR model, the exponential model, the free-parton model,
and the Gagenbauer-1(2) models are of similar magnitude,
whereas the numbers from the Gaussian-type  are smaller by about 1 order of magnitude.
The main reason is that the Gaussian-type model yields a severe suppression at the end point region.
The shape parameters in the LCDAs of the $\Lambda$ baryon in the literature vary drastically,
which means that our form factor results were strongly sensitive to  different parameter sets.
It has been demonstrated that the COZ models may be the most suitable choices to
obtain a reasonable estimate under the PQCD  formulism within the accuracy of the current work.
The obtained two baryonic transition form factors $f_1$ and $g_1$ at maximum recoil
are in accord with the expectation in the heavy-quark limit
and agree well with the existing results in other works within errors.

\item[(II)]
We have computed four independent complex helicity amplitudes allowed by angular momentum conservation,
which are linearly related to the invariant amplitudes.
It has been observed that negative-helicity amplitudes dominate,
and the relative contributions from the positive ones only amount to a few percent.
In PQCD formulism, the nonfactorizable contributions and vertex corrections provide the main sources of the strong phase.
The weak phases from the related CKM matrix elements are almost zero to order $\lambda^2$; thus, no direct $CP$ violation is expected.
Setting one of the phases to be zero, another three relative phases have been estimated,
and the obtained values fall into the $95\%$ credibility interval reported by LHCb.

\item[(III)]
With the helicity amplitude results at hand, we can compute some interesting observables such as branching ratios
as well as the various measurable asymmetries.
Our prediction $\mathcal{B}(\Lambda_b\rightarrow \Lambda J/\psi)=7.75_{-3.17-0.96}^{+4.08+1.39}$ coincides with the CCQM calculation,
but $\mathcal{B}(\Lambda_b\rightarrow \Lambda \psi(2S))$ is only half of them.
The ratio of the two branching ratios is predicted to be $0.47^{+0.02}_{-0.00}$,
which is smaller than those from the other approaches and more consistent with the latest average $0.508\pm 0.023$.
The predicted up-down asymmetries $\alpha_b$ for the two modes in PQCD differ a lot
due to the  significant nonfactorizable contributions, which are process dependent.
Our number for the $J/\psi$ mode is closer to the most probable value reported by LHCb.
The predicted asymmetry for the $\psi(2S)$ mode is large in size
and can be tested in future experiments.
Moreover, we have also calculated the various observable parameters such as $\alpha_2$, $r_{0,1}$, and $\gamma_0$
defined in terms of the linear combinations of normalized squared helicity amplitudes.
The obtained results
are compatible within uncertainties with the CCQM calculations and the measurement from the LHCb Collaboration.
\item[(IV)]
We have discussed discuss the long-distance effects, arising from the charmonium resonances regions
in the semileptonic decay $\Lambda_b\rightarrow \Lambda l^+l^-$  by using the zero width approximation.
Our result for the muonic mode is comparable with other predictions,
but for the tauonic one, the various predictions including ours have an obvious discrepancy,
which should be clarified in the future.
\end{itemize}
In brief,  the PQCD formalism could be well applied to the concerned color-suppressed $\Lambda_b$ decays,
although our theoretical predictions are still plagued by larger uncertainties due to the hadronic parameters.
Some of the obtained observables are compatible with the current data and other theoretical predictions,
while others could be measured in the ongoing experiments with certain precision.

\begin{acknowledgments}
We thank Hsiang-nan Li, Fu-Sheng Yu, Yue-Long Shen, and Ya Li for helpful discussions.
This work is supported by National Natural Science Foundation
of China under Grants No. 12075086 and  No. 11605060 and the Natural Science Foundation of Hebei Province
under Grants No.A2021209002 and  No.A2019209449.
\end{acknowledgments}

\begin{appendix}
\section{VARIOUS LCDA MODELS OF $\Lambda_b$ BARYON}\label{sec:LCDAs}
We collect the five widely used parametrized models  for the LCDAs of the $\Lambda_b$ baryon as follows:
\begin{itemize}
\item[$\bullet$]Exponential model~\cite{jhep112013191}
\begin{eqnarray}
\Psi_2(x_2,x_3)&=&     x_2x_3\frac{M^4}{\omega_0^4}e^{-\frac{\omega}{\omega_0}},\nonumber\\
\Psi_3^{+-}(x_2,x_3)&=&2x_2\frac{M^3}  {\omega_0^3}e^{-\frac{\omega}{\omega_0}},\nonumber\\
\Psi_3^{-+}(x_2,x_3)&=&2x_3\frac{M^3}  {\omega_0^3}e^{-\frac{\omega}{\omega_0}},\nonumber\\
\Psi_4(x_2,x_3)&=&     \frac{ M^2}     {\omega_0^2}e^{-\frac{\omega}{\omega_0}},
\end{eqnarray}
with $\omega=(x_2+x_3)M$ and $\omega_0=0.4\pm0.1$ GeV.

\item[$\bullet$] Free-parton approximation~\cite{jhep112013191}
\begin{eqnarray}
\Psi_2(x_2,x_3)&=&     \frac{15x_2x_3M^4(2\bar{\Lambda}-\omega)  }{4\bar{\Lambda}^5}\Theta(2\bar{\Lambda}-\omega),\nonumber\\
\Psi_3^{+-}(x_2,x_3)&=&\frac{15x_2   M^3(2\bar{\Lambda}-\omega)^2}{4\bar{\Lambda}^5}\Theta(2\bar{\Lambda}-\omega),\nonumber\\
\Psi_3^{-+}(x_2,x_3)&=&\frac{15x_3   M^3(2\bar{\Lambda}-\omega)^2}{4\bar{\Lambda}^5}\Theta(2\bar{\Lambda}-\omega),\nonumber\\
\Psi_4(x_2,x_3)&=&     \frac{5       M^2(2\bar{\Lambda}-\omega)^3}{8\bar{\Lambda}^5}\Theta(2\bar{\Lambda}-\omega),
\end{eqnarray}
with $\bar{\Lambda} \equiv (M-m_b)/2 \approx 0.8$ GeV.

\item[$\bullet$] Gegenbauer-1 model~\cite{plb665197}
\begin{eqnarray}
\Psi_2(x_2,x_3)&=&     M^4x_2x_3   \left[\frac{1}{\varepsilon_0^4}e^{-\frac{\omega}{\varepsilon_0}}+
\frac{a_2}{\varepsilon_1^4}C_2^{3/2}\left(\frac{x_2-x_3}{x_2+x_3}\right)e^{-\frac{\omega}{\varepsilon_1}}\right],\nonumber\\
\Psi_3^{+-}(x_2,x_3)&=&2M^3x_2\frac{1}{\varepsilon_3^3}e^{-\frac{\omega}{\varepsilon_3}},\nonumber\\
\Psi_3^{-+}(x_2,x_3)&=&2M^3x_3\frac{1}{\varepsilon_3^3}e^{-\frac{\omega}{\varepsilon_3}},\nonumber\\
\Psi_4(x_2,x_3)&=&     5M^2\mathcal{N}^{-1}\int_{\frac{\omega}{2}}^{s_0}dse^{-s/\tau}\left(s-\frac{\omega}{2}\right)^3.
\end{eqnarray}
with the constant $\mathcal{N}=\int_{0}^{s_0}dss^5e^{-s/\tau}$ and other parameters $0.4<\tau<0.8$ GeV, $s_0=1.2$ GeV, $a_2=0.333_{-0.333}^{+0.250}$, $\varepsilon_0=200_{-60}^{+130}$ MeV, $\varepsilon_1=650_{-300}^{+650}$ MeV, and $\varepsilon_3=230$ MeV.

\item[$\bullet$]  Gegenbauer-2 model~\cite{Ali:2012zza}
\begin{eqnarray}
\Psi_2(x_2,x_3)&=&     M^4x_2x_3    \sum_{n=0}^2\frac{a_n^{(2)}}{{\varepsilon_n^{(2)}}^4}C_n^{3/2}\left(\frac{x_2-x_3}{x_2+x_3}\right)e^{-\frac{\omega}{\varepsilon_n^{(2)}}},\nonumber\\
\Psi_3^{+-}(x_2,x_3)&=&M^3(x_2+x_3)\left[\sum_{n=0}^2\frac{a_n^{(3)}}{{\varepsilon_n^{(3)}}^3}C_n^{1/2}\left(\frac{x_2-x_3}{x_2+x_3}\right)
e^{-\frac{\omega}{\varepsilon_n^{(3)}}}
                       +            \sum_{n=0}^3\frac{b_n^{(3)}}{{\eta       _n^{(3)}}^3}C_n^{1/2}\left(\frac{x_2-x_3}{x_2+x_3}\right)e^{-\frac{\omega}{\eta_n^{(3)}}}\right],\nonumber\\
\Psi_3^{-+}(x_2,x_3)&=&M^3(x_2+x_3)\left[\sum_{n=0}^2\frac{a_n^{(3)}}{{\varepsilon_n^{(3)}}^3}C_n^{1/2}\left(\frac{x_2-x_3}{x_2+x_3}\right)
e^{-\frac{\omega}{\varepsilon_n^{(3)}}}
                       -            \sum_{n=0}^3\frac{b_n^{(3)}}{{\eta       _n^{(3)}}^3}C_n^{1/2}\left(\frac{x_2-x_3}{x_2+x_3}\right)e^{-\frac{\omega}{\eta_n^{(3)}}}\right],\nonumber\\
\Psi_4(x_2,x_3)&=&     M^2          \sum_{n=0}^2\frac{a_n^{(4)}}{{\varepsilon_n^{(4)}}^2}C_n^{1/2}\left(\frac{x_2-x_3}{x_2+x_3}\right)e^{-\frac{\omega}{\varepsilon_n^{(4)}}},
\end{eqnarray}
with the Gegenbauer polynomials
\begin{eqnarray}
C_0^{\xi}(x)&=&1, \quad C_1^{\xi}(x)=2\xi x, \quad C_2^{\xi}(x)=2\xi(1+\xi)x^2-\xi. 
\end{eqnarray}
The relevant parameters for the Gegenbauer-2 model 
can be found in~\cite{Ali:2012zza}.

\item[$\bullet$]  QCDSR model~\cite{plb665197}
\begin{eqnarray}
\Psi_2(x_2,x_3)&=&     \frac{15}{2\mathcal{N}}M^4x_2x_3 \int_{\frac{\omega}{2}}^{s_0}dse^{-s/\tau}(s-\frac{\omega}{2})  ,\nonumber\\
\Psi_3^{+-}(x_2,x_3)&=&\frac{15}{\mathcal{N}}M^3x_2     \int_{\frac{\omega}{2}}^{s_0}dse^{-s/\tau}(s-\frac{\omega}{2})^2,\nonumber\\
\Psi_3^{-+}(x_2,x_3)&=&\frac{15}{\mathcal{N}}M^3x_3     \int_{\frac{\omega}{2}}^{s_0}dse^{-s/\tau}(s-\frac{\omega}{2})^2,\nonumber\\
\Psi_4(x_2,x_3)&=&     \frac{5}{\mathcal{N}}M^2         \int_{\frac{\omega}{2}}^{s_0}dse^{-s/\tau}(s-\frac{\omega}{2})^3.
\end{eqnarray}

The various models above obey the uniform normalization conditions according to the different twists,
\begin{eqnarray}
&&\int_0^1dx_2dx_3\Psi_2(x_2,x_3)=1,\nonumber\\
&&\int_0^1dx_2dx_3(\Psi_3^{+-}(x_2,x_3)+\Psi_3^{-+}(x_2,x_3))/4=1,\nonumber\\
&&\int_0^1dx_2dx_3\Psi_4(x_2,x_3)=1.
\end{eqnarray}

\item[$\bullet$]   Gaussian-type~\cite{prd531416,zpc51321} \\
This model does not distinguish the asymptotic forms of different twists' LCDAs,
in favor of an unifying Gaussian shape~\cite{9211255}
\begin{eqnarray}
\Phi_{\Lambda_b}(x_1,x_2,x_3)=Nx_1x_2x_3\exp[-\frac{1}{2\beta^2}(\frac{M^2}{x_1}+\frac{m_q^2}{x_2}+\frac{m_q^2}{x_3})],
\end{eqnarray}
with the normalized constant $N=6.67\times10^{12}$ and shape parameters $\beta=1.0$ GeV and $m_q=0.3$ GeV.
The  corresponding  $\Lambda_b$ baryon LCDA  can be expressed as
\begin{eqnarray}
(\Psi_{\Lambda_b})_{\alpha\beta\gamma}=\frac{f_{\Lambda_b}}{8\sqrt{2}N_c}[(\rlap{/}{p}+M)\gamma_5 C]_{\beta\gamma}[\Lambda_b(p)]_\alpha
\Phi_{\Lambda_b}(x_1,x_2,x_3).
\end{eqnarray}

\end{itemize}
\section{FACTORIZATION FORMULAS}\label{sec:app}
\begin{table}[!htbh]
\caption{The expressions of $V^{LL}$ and $V^{SP}$ in Eq.~(\ref{eq:amp}) for each diagram $T_{ij}$.}
\newcommand{\tabincell}[2]{\begin{tabular}{@{}#1@{}}#2\end{tabular}}
\label{tab:wilson}
\begin{tabular}[t]{lcc}
\hline\hline
$ij$                   &$V^{LL}$                       &$V^{SP}$             \\ \hline
$a1,a2,a3,a5,b1,b2,b4$       &$V_{cb}V^*_{cs}a_1+V_{tb}V^*_{ts}(a_3+a_9)$    &$V_{tb}V^*_{ts}(a_5+a_7)$          \\
$a6,a7,b6,b7,c1,c2,d1,d2$  &$\frac{1}{3}V_{cb}V^*_{cs}C_2+\frac{1}{3}V_{tb}V^*_{ts}(C_4+C_{10})$
             &$\frac{1}{3}V_{tb}V^*_{ts}(C_6+C_8)$                      \\
$c5,c7,d6$ &$V_{cb}V^*_{cs}(\frac{1}{3}C_2-\frac{1}{4}C_1)+V_{tb}V^*_{ts}[\frac{1}{3}(C_4+C_{10})-\frac{1}{4}(C_3+C_9)]$
&$V_{tb}V^*_{ts}[\frac{1}{3}(C_6+C_8)-\frac{1}{4}(C_5+C_7)]$ \\
\hline\hline
\end{tabular}
\end{table}
Following the conventions in Ref.~\cite{220209181},
we provide some details about the factorization formulas in Eq.~(\ref{eq:amp}).
The formulas of the equivalent diagrams connected by an
interchange of two light quarks will not be repeated here.
The explicit expressions of $V^{LL,SP}$ are collected in Table~\ref{tab:wilson},
where the combinations of the Wilson coefficients $a_l$ including the vertex corrections are defined as
\begin{align}
\begin{split}
a_l=\left \{
\begin{array}{ll}
C_l+\frac{1}{3}C_{l+1}+\frac{\alpha_s}{4\pi}\frac{C_f}{N_c}C_{l+1}[-18-12ln(\frac{t}{m_b})+f_I], &  l=1,3,9, \\
C_l+\frac{1}{3}C_{l+1}-\frac{\alpha_s}{4\pi}\frac{C_f}{N_c}C_{l+1}[-6-12ln(\frac{t}{m_b})+f_I],  &  l=5,7,
\end{array}
\right.
\end{split}
\end{align}
with $C_f=(N^2_c-1)/(2N_c)$.
For the calculation of $f_I$, the reader is referred to Refs.~\cite{Cheng:2000kt,Cheng:2001ez} for details.

The virtualities of the internal propagators $t_{A,B,C,D}$ and the hard amplitudes $H_{T_{ij}}$ for each diagram $T_{ij}$ in Fig.~\ref{fig:Feynman} are gathered in the Tables~\ref{tab:ttt} and~\ref{tab:amp}, respectively.
The expression of $H_{T_{ij}}$ in Table~\ref{tab:amp} is given for $A_1^L$ and $A_2^L$,
where those terms proportional to $r_{\Lambda}$ have been neglected for simplicity.
The corresponding formulas for $B_1^L$ and $B_2^L$ can be obtained from $A_1^L$ and $A_2^L$, respectively,
by the following replacement:
\begin{eqnarray}
B_1^L=A_1^L|_{\Phi^{V}\rightarrow -\Phi^{V},\Phi^{A}\rightarrow -\Phi^{A}}, \quad
B_2^L=A_2^L|_{\Phi^{T}\rightarrow -\Phi^{T}}.
\end{eqnarray}
The formulas of $A_{1}^T(B_{1}^T)$  have the same form as $A_{1}^L(B_{1}^L)$
but with $\psi^L\rightarrow \psi^V$ and $\psi^t\rightarrow \psi^T$.
\begin{table}[!htbh]
\caption{The virtualities of the internal gluon $t_{A,B}$ and quark $t_{C,D}$ for each diagram $T_{ij}$ in Fig.~\ref{fig:Feynman}. }
\newcommand{\tabincell}[2]{\begin{tabular}{@{}#1@{}}#2\end{tabular}}
\label{tab:ttt}
\begin{tabular}[t]{lcccc}
\hline\hline
$T_{ij}$&$\frac{t_A}{M^2}$&$\frac{t_B}{M^2}$   &$\frac{t_C}{M^2}$   &$\frac{t_D}{M^2}$\\ \hline
$T_{a1}$&$f^+x_3x_3'$     &$f^+(1-x_1)(1-x_1')$&$f^+(1-x_1)x_3'$    &$f^+(1-x_1')$\\
$T_{a2}$&$f^+x_3x_3'$     &$f^+(1-x_1)(1-x_1')$&$f^+(1-x_1')x_3$    &$f^+(1-x_1')$\\
$T_{a3}$&$f^+x_3x_3'$     &$f^+x_2x_2'$        &$x_2+f^+(1-x_2)x_3'$&$f^+(1-x_1')$\\
$T_{a5}$&$f^+x_3x_3'$     &$f^+x_2x_2'$        &$f^+(1-x_2')x_3$    &$x_3+f^+(1-x_3)x_2'$\\
$T_{a6}$&$f^+x_3x_3'$     &$f^+x_2x_2'$        &\tabincell{c}{$r_c^2+f^+x_3x_3'+r_{\Lambda}^2x_3'y$\\$-f^+(x_3+x_3')y+y(x_3-r^2y)$}     &$x_3+f^+(1-x_3)x_2'$\\
$T_{a7}$&$f^+x_3x_3'$     &$f^+x_2x_2'$        &\tabincell{c}{$r_c^2+f^+x_3x_3'+((f^+-1)x_3$\\$+(f^+-r_{\Lambda}^2)x_3'-r^2(y-1))(y-1)$}&$x_3+f^+(1-x_3)x_2'$\\
$T_{b1}$&$f^+x_3x_3'$     &$f^+(1-x_1)(1-x_1')$&$f^+(1-x_1)x_3'$    &$f^+(1-x_1)$\\
$T_{b2}$&$f^+x_3x_3'$     &$f^+(1-x_1)(1-x_1')$&$f^+(1-x_1')x_3$    &$f^+(1-x_1)$\\
$T_{b4}$&$f^+x_3x_3'$     &$f^+x_2x_2'$        &$f^+(1-x_1)$        &$f^+(1-x_3')x_2$\\
$T_{b6}$&$f^+x_3x_3'$     &$f^+x_2x_2'$        &\tabincell{c}{$r_c^2+f^+x_3x_3'+r_{\Lambda}^2x_3'y$\\$-f^+(x_3+x_3')y+y(x_3-r^2y)$}     &$f^+(1-x_3')x_2$\\
$T_{b7}$&$f^+x_3x_3'$     &$f^+x_2x_2'$        &\tabincell{c}{$r_c^2+f^+x_3x_3'+((f^+-1)x_3$\\$+(f^+-r_{\Lambda}^2)x_3'-r^2(y-1))(y-1)$}&$f^+(1-x_3')x_2$\\
$T_{c1}$&$f^+x_3x_3'$     &$f^+(1-x_1)(1-x_1')$&$f^+(1-x_1)x_3'$    &\tabincell{c}{$r_c^2+f^+y(x_1+x_1'-2)+y(1-r^2y-x_1)$\\$+(f^+(x_1-1)-yr_{\Lambda}^2)(x_1'-1)$}\\
$T_{c2}$&$f^+x_3x_3'$     &$f^+(1-x_1)(1-x_1')$&$f^+(1-x_1')x_3$    &\tabincell{c}{$r_c^2+f^+y(x_1+x_1'-2)+y(1-r^2y-x_1)$\\$+(f^+(x_1-1)-yr_{\Lambda}^2)(x_1'-1)$}\\
$T_{c5}$&$f^+x_3x_3'$     &$f^+x_2x_2'$        &\tabincell{c}{$r_c^2+f^+x_3x_3'+r_{\Lambda}^2x_3'y$\\$-f^+(x_3+x_3')y+y(x_3-r^2y)$}
&\tabincell{c}{$r_c^2+f^+y(x_1+x_1'-2)+y(1-r^2y-x_1)$\\$+(f^+(x_1-1)-yr_{\Lambda}^2)(x_1'-1)$}\\
$T_{c7}$&$f^+x_3x_3'$     &$f^+x_2x_2'$        &\tabincell{c}{$r_c^2+f^+x_3x_3'+((f^+-1)x_3$\\$+(f^+-r_{\Lambda}^2)x_3'-r^2(y-1))(y-1)$}
&\tabincell{c}{$r_c^2+f^+x_2x_2'+r_{\Lambda}^2x_2'y$\\$-f^+(x_2+x_2')y+y(x_2-r^2y)$}\\
$T_{d1}$&$f^+x_3x_3'$     &$f^+(1-x_1)(1-x_1')$&$f^+(1-x_1)x_3'$    &\tabincell{c}{$r_c^2-f^+(y(x_1+x_1'-2)-x_1-x_1'+1)$\\$(y-1)(r^2(y-1)-x_1+1-r_{\Lambda}^2(1-x_1'))$}\\
$T_{d2}$&$f^+x_3x_3'$     &$f^+(1-x_1)(1-x_1')$&$f^+(1-x_1')x_3$    &\tabincell{c}{$r_c^2-f^+(y(x_1+x_1'-2)-x_1-x_1'+1)$\\$(y-1)(r^2(y-1)-x_1+1-r_{\Lambda}^2(1-x_1'))$}\\
$T_{d6}$&$f^+x_3x_3'$     &$f^+x_2x_2'$        &\tabincell{c}{$r_c^2-f^+(y(x_1+x_1'-2)-x_1-x_1'+1)$\\$(y-1)(r^2(y-1)-x_1+1-r_{\Lambda}^2(1-x_1'))$}
&\tabincell{c}{$r_c^2+f^+x_3x_3'+((f^+-1)x_3$\\$+(f^+-r_{\Lambda}^2)x_3'-r^2(y-1))(y-1)$}\\
\hline\hline
\end{tabular}
\end{table}

\begin{table}[H]
\centering
\caption{The expressions of $H^{LL,SP}_{R_{ij}}$ for $A_1^L$ and $A_2^L$.}
\newcommand{\tabincell}[2]{\begin{tabular}{@{}#1@{}}#2\end{tabular}}
\label{tab:amp}
\begin{tabular}[t]{lcc}
\hline\hline
$$                   &$\frac{A_1^L}{16M^4}$&$\frac{A_2^L}{16M^4}$\\ \hline
$H_{T_{a1}}^{LL(SP)}$&\tabincell{c}{$[-2r(1-r^2)^2x_3'\Psi_4\psi^L]\Phi^A$}
                     &\tabincell{c}{$[2r(1-r^2)(x_1-1)(x_1'-1)\psi^L(\Psi_3^{-+}+\Psi_3^{+-})]\Phi^T$}\\
$H_{T_{a2}}^{LL(SP)}$&\tabincell{c}{$[2r(r^2-1)x_3\Psi_4\psi^L]\Phi^A$}
                     &\tabincell{c}{$[4r(1-r^2)x_3(x_1'-1)\psi^L\Psi_4]\Phi^A$}\\
$H_{T_{a3}}^{LL(SP)}$&\tabincell{c}{$[r(r^2-1)((r^2-1)x_3'-x_2)\Psi_4\psi^L]\Phi^V$\\$ + [r(r^2-1)(((r^2-1)x_3'$\\$-x_2)\Psi_4+2(1-x_2)\Psi_2)\psi^L]\Phi^A$}
                     &\tabincell{c}{$[2r(2(1-r^2)x_3'-x_2(r^2(x_1'-1)-x_1'-1))\psi^L\Psi_4]\Phi^V$\\$ + [2r(r^2-1)x_2(x_1'-1)\psi^L\Psi_4]\Phi^A$\\$ + [2r(((1-x_2)(r^2(x_1'-1)-x_1')+1)\Psi_3^{-+}+x_2\Psi_3^{+-})\psi^L]\Phi^T$}\\
$H_{T_{a5}}^{LL(SP)}$&\tabincell{c}{$[r(r^2-1)x_3\Psi_3^{+-}\psi^L]\Phi^T$}
                     &\tabincell{c}{$[-2r^3x_3^2\psi^L\Psi_4]\Phi^V + [-2r^3x_3^2\psi^L\Psi_4]\Phi^A$\\$ + [-2rx_3(r^2(x_3-1)+1)\psi^L\Psi_3^{+-}]\Phi^T$}\\
$H_{T_{a6}}^{LL}$    &\tabincell{c}{$[2r(r^2-1)(x_3-1)(x_3-y)\psi^L\Psi_2]\Phi^A$\\$ + [(1-r^2)(r_c\psi^t+r\psi^L(x_3-y))\Psi_3^{+-}]\Phi^T$}
                     &\tabincell{c}{$[2r^2x_3(r_c\psi^t+r\psi^L(x_3-y))\Psi_4]\Phi^V$\\$ + [2rx_3(rr_c\psi^t\Psi_4+(r^2\Psi_4-2\Psi_2)(x_3-y)\psi^L)]\Phi^A$\\$ + [2(r^2(x_3-1)+1)(r_c\psi^t+r\psi^L(x_3-y))\Psi_3^{+-}$\\$+2r(r^2-1)(x_3-y)x_2'\psi^L\Psi_3^{-+}]\Phi^T$}\\
$H_{T_{a6}}^{SP}$    &\tabincell{c}{$[-(1-r^2)^2x_2'(r_c\psi^t+r\psi^L(x_3-y))\Psi_4]\Phi^V$\\$ + [(r^2-1)^2x_2'(r_c\psi^t+r\psi^L(x_3$\\$-y))\Psi_4-2(r^2-1)(x_3-1)r_c\psi^t\Psi_2]\Phi^A$}
                     &\tabincell{c}{$[-2(r_c\psi^t(r^2(x_3-x_2')+x_2')+r\psi^L(r^2y(x_2'-x_3)$\\$+(r^2-1)x_3(x_3'-x_2')-yx_2'))\Psi_4]\Phi^V$\\$ + [4((x_3-1)r_c\psi^t+r\psi^L(y-x_3))\Psi_2$\\$+2(r\psi^L((r^2-1)yx_2'+r^2x_3y-(r^2-1)x_3(x_2'+x_3'))$\\$-r_c\psi^t(r^2(x_2'+x_3)-x_2'))\Psi_4]\Phi^A$\\$ + [-2r((x_3-1)(rr_c\psi^t+\psi^L((r^2-1)x_3'-r^2y))\Psi_3^{+-}$\\$-(r^2-1)(x_3-y)x_2'\psi^L\Psi_3^{-+})]\Phi^T$}\\
$H_{T_{a7}}^{LL}$    &\tabincell{c}{$[(r^2-1)^2x_2'(r_c\psi^t+r\psi^L(x_3+y-1))\Psi_4]\Phi^V$\\$ + [-(1-r^2)^2x_2'(r_c\psi^t+r\psi^L(x_3$\\$+y-1))\Psi_4+2(r^2-1)(x_3-1)r_c\psi^t\Psi_2]\Phi^A$}
                     &\tabincell{c}{$[2(r_c\psi^t(r^2(x_3-x_2')+x_2')-r\psi^L((r^2$\\$-1)(x_3+y-1)x_2'+x_3(x_3'-r^2(x_3'+y-1))))\Psi_4]\Phi^V$\\$ + [2(r_c(r^2(x_2'+x_3)-x_2')\psi^t+r(r^2x_3(x_3'+y$\\$-1)+(r^2-1)(x_3+y-1)x_2'-x_3x_3')\psi^L)\Psi_4$\\$+4(r\psi^L(x_3+y-1)-(x_3-1)r_c\psi^t)\Psi_2]\Phi^A$\\$ + [2r((x_3-1)(rr_c\psi^t+\psi^L(r^2(x_3'+y-1)-x_3'))\Psi_3^{+-}$\\$-(r^2-1)(x_3+y-1)x_2'\psi^L\Psi_3^{-+})]\Phi^T$}\\
$H_{T_{a7}}^{SP}$    &\tabincell{c}{$[2r(r^2-1)(1-x_3)(x_3+y-1)\psi^L\Psi_2]\Phi^A$\\$ + [(r^2-1)(r_c\psi^t+r\psi^L(x_3+y-1))\Psi_3^{+-}]\Phi^T$}
                     &\tabincell{c}{$[-2r^2x_3(r_c\psi^t+r\psi^L(x_3+y-1))\Psi_4]\Phi^V$\\$ + [2rx_3(\psi^L(2\Psi_2-r^2\Psi_4)(x_3+y-1)-rr_c\psi^t\Psi_4)]\Phi^A$\\$ + [2r(1-r^2)(x_3+y-1)x_2'\psi^L\Psi_3^{-+}-2(r^2(x_3$\\$-1)+1)(r_c\psi^t+r\psi^L(x_3+y-1))\Psi_3^{+-}]\Phi^T$}\\
$H_{T_{b1}}^{LL(SP)}$&\tabincell{c}{$[r(1-r^2)(x_1-1)^2(\Psi_3^{-+}+\Psi_3^{+-})\psi^L]\Phi^T$}
                     &\tabincell{c}{$[2r(1-r^2)(x_1-1)^2\psi^L(\Psi_3^{-+}+\Psi_3^{+-})]\Phi^T$}\\
$H_{T_{b2}}^{LL(SP)}$&\tabincell{c}{$[-2r(1-r^2)^2x_3\Psi_4\psi^L]\Phi^A$}
                     &\tabincell{c}{$0$}\\
$H_{T_{b4}}^{LL(SP)}$&\tabincell{c}{$[-r(1-r^2)^2x_2\Psi_4\psi^L]\Phi^V$\\$ + [r(r^2-1)^2x_2\Psi_4\psi^L]\Phi^A$\\$ + [r(r^2-1)(1-x_1)x_2\Psi_3^{-+}\psi^L]\Phi^T$}
                     &\tabincell{c}{$[2r(r^2-1)(1-x_1)x_2\psi^L\Psi_3^{-+}]\Phi^T$}\\
$H_{T_{b6}}^{LL}$    &\tabincell{c}{$[r(r^2-1)x_2(rr_c\psi^t+\psi^L((r^2-1)x_3'-r^2y))\Psi_4]\Phi^V$\\$ + [r(1-r^2)x_2(rr_c\psi^t+\psi^L((r^2-1)x_3'-r^2y))\Psi_4]\Phi^A$\\$ + [(1-r^2)x_2r_c\psi^t\Psi_3^{-+}]\Phi^T$}
                     &\tabincell{c}{$[2(1-r^2)x_2r_c\psi^t\Psi_3^{-+}]\Phi^T$}\\
$H_{T_{b6}}^{SP}$    &\tabincell{c}{$[r(r^2-1)x_2(x_3-y)\psi^L\Psi_3^{-+}]\Phi^T$}
                     &\tabincell{c}{$[-2r^2x_2(r_c\psi^t+r\psi^L(x_3-y))\Psi_4]\Phi^V$\\$ + [2r^2x_2(r_c\psi^t+r\psi^L(x_3-y))\Psi_4]\Phi^A$\\$ + [-2rx_2(rr_c\psi^t+\psi^L(x_3-y))\Psi_3^{-+}]\Phi^T$}\\
$H_{T_{b7}}^{LL}$    &\tabincell{c}{$[r(1-r^2)x_2(x_3+y-1)\psi^L\Psi_3^{-+}]\Phi^T$}
                     &\tabincell{c}{$[2r^2x_2(r_c\psi^t+r\psi^L(x_3+y-1))\Psi_4]\Phi^V$\\$ + [-2r^2x_2(r_c\psi^t+r\psi^L(x_3+y-1))\Psi_4]\Phi^A$\\$ + [2rx_2(rr_c\psi^t+\psi^L(x_3+y-1))\Psi_3^{-+}]\Phi^T$}\\
$H_{T_{b7}}^{SP}$    &\tabincell{c}{$[r(1-r^2)x_2(rr_c\psi^t+\psi^L(r^2(x_3'+y-1)-x_3'))\Psi_4]\Phi^V$\\$ + [r(r^2-1)x_2(rr_c\psi^t+\psi^L(r^2(x_3'+y-1)-x_3'))\Psi_4]\Phi^A$\\$ + [(r^2-1)x_2r_c\psi^t\Psi_3^{-+}]\Phi^T$}
                     &\tabincell{c}{$[2(r^2-1)x_2r_c\psi^t\Psi_3^{-+}]\Phi^T$}\\
$H_{T_{c1}}^{LL}$    &\tabincell{c}{$[-2r(1-r^2)^2(x_1+y-1)x_3'\psi^L\Psi_4]\Phi^A$\\$ + [(r^2-1)(x_1-1)(r\psi^L(x_1$\\$+y-1)-r_c\psi^t)(\Psi_3^{-+}+\Psi_3^{+-})]\Phi^T$}
                     &\tabincell{c}{$[2(r^2-1)(x_1-1)(r\psi^L(x_1$\\$+y-1)-r_c\psi^t)(\Psi_3^{-+}+\Psi_3^{+-})]\Phi^T$}\\
$H_{T_{c1}}^{SP}$    &\tabincell{c}{$[-2(1-r^2)^2r_cx_3'\psi^t\Psi_4]\Phi^A$}
                     &\tabincell{c}{$[4(1-r^2)x_3'(r\psi^L(x_1+y-1)-r_c\psi^t)\Psi_4]\Phi^A$\\$ + [2r(1-x_1)(\psi^L(r^2(x_1'+y-1)$\\$-x_1'+1)-rr_c\psi^t)(\Psi_3^{-+}+\Psi_3^{+-})]\Phi^T$}\\
\hline\hline
\end{tabular}
\end{table}

\begin{table}[H]
\centering
TABLE~\ref{tab:amp} (continued)
\newcommand{\tabincell}[2]{\begin{tabular}{@{}#1@{}}#2\end{tabular}}
\label{tab:amp2}
\begin{tabular}[t]{lcc}
\hline\hline
$$                  &$\frac{A_1^L}{16M^4}$&$\frac{A_2^L}{16M^4}$\\ \hline
$H_{T_{c2}}^{LL}$   &\tabincell{c}{$[2r^2(r^2-1)x_3r_c\psi^t\Psi_4]\Phi^A$}
                    &\tabincell{c}{$0$}\\
$H_{T_{c2}}^{SP}$   &\tabincell{c}{$[2r(r^2-1)x_3(r^2(x_1'+y-1)-x_1'+1)\psi^L\Psi_4]\Phi^A$}
                    &\tabincell{c}{$[-4rx_3(\psi^L(r^2(x_1'+y-1)-x_1'+1)-rr_c\psi^t)\Psi_4]\Phi^A$}\\
$H_{T_{c5}}^{LL}$   &\tabincell{c}{$[r(r^2-1)((r_c^2-(x_1+y-1)((1$\\$-r^2)x_3'+r^2y))\psi^L-rr_cx_2\psi^t)\Psi_4]\Phi^V$\\$ + [r(1-r^2)((\psi^L(r_c^2-(x_1+y-1)((1-r^2)x_3'+r^2y))$\\$-rr_cx_2\psi^t)\Psi_4+2(r^2-1)(x_1+y-1)(x_3-y)\psi^L\Psi_2)]\Phi^A$\\$ + [(r^2-1)(1-x_1-y)(r_c\psi^t+r\psi^L(x_3$\\$-y))\Psi_3^{+-}+(r^2-1)r_cx_2\psi^t\Psi_3^{-+}]\Phi^T$}
                    &\tabincell{c}{$[2(1-r^2)((x_1+y-1)(r_c\psi^t$\\$+r\psi^L(x_3-y))\Psi_3^{+-}-r_cx_2\psi^t\Psi_3^{-+})]\Phi^T$}\\
$H_{T_{c5}}^{SP}$   &\tabincell{c}{$[(r(r^2-1)\psi^L(r_c^2+(x_3-y)(r^2(x_1'+y$\\$-1)-x_1'+1))-(r^2-1)r_cx_2'\psi^t)\Psi_4]\Phi^V$\\$ + [(1-r^2)((r\psi^L(r_c^2+(x_3-y)(r^2(x_1'+y-1)-x_1'$\\$+1))-(r^2-1)r_c\psi^tx_2')\Psi_4+2(r^2-1)r_c(x_3-y)\psi^t\Psi_2)]\Phi^A$}
                    &\tabincell{c}{$[-2(r\psi^L(2r_c^2-(r^2-1)(x_3-y)x_2'+(x_2+2x_3$\\$-2y)((1-r^2)x_3'+r^2y))+r_c\psi^t(x_2'-r^2(x_2'+x_2)))\Psi_4]\Phi^V$\\$ + [2(r_c\psi^t(r^2(x_2-x_2')+x_2')$\\$-r\psi^L(r^2x_2(y-x_3')+(r^2-1)(x_3-y)x_2'+x_2x_3'))\Psi_4$\\$+4(x_3-y)(r_c\psi^t+r\psi^L(x_2+x_3-y))\Psi_2]\Phi^A$\\$ + [-2r((r_c^2+(x_3-y)(r^2(x_1'+y-1)-x_1'$\\$+1))\psi^L\Psi_3^{-+}+r_c(r_c\psi^L+r\psi^t(x_3-y))\Psi_3^{+-})]\Phi^T$}\\
$H_{T_{c7}}^{LL}$   &\tabincell{c}{$[2(r^2-1)r_c(x_2-y)\psi^t\Psi_2]\Phi^A$\\$ + [(r^2-1)(1-x_3-y)(r\psi^L(y-x_2)-r_c\psi^t)\Psi_3^{-+}]\Phi^T$}
                    &\tabincell{c}{$[2r\Psi_4(\psi^L(r^2(x_3+y-1)(y-x_2)-r_c^2)+rx_1r_c\psi^t)]\Phi^V$\\$ + [2r(\psi^L(r_c^2+r^2(x_2-y)(x_3+y-1))-rx_1r_c\psi^t)\Psi_4$\\$-4(x_2-y)(r_c\psi^t+r\psi^L(x_3+y-1))\Psi_2]\Phi^A$\\$ + [2r(x_2-y)(rr_c\psi^t+\psi^L(r^2(x_3'+y-1)-x_3'))\Psi_3^{+-}$\\$-2(r\psi^L(r_c^2+(x_2-y)(x_3+y-1))$\\$+r_c\psi^t(r^2(x_2-y)+x_3+y-1))\Psi_3^{-+}]\Phi^T$}\\
$H_{T_{c7}}^{SP}$   &\tabincell{c}{$[2(r^2-1)r_c(x_3+y-1)\psi^t\Psi_2]\Phi^A$\\$ + [(r^2-1)(y-x_2)(r_c\psi^t+r\psi^L(x_3+y-1))\Psi_3^{+-}]\Phi^T$}
                    &\tabincell{c}{$[-2r\Psi_4(\psi^L(r^2(x_3+y-1)(y-x_2)-r_c^2)+rx_1r_c\psi^t)]\Phi^V$\\$ + [-2(r(\psi^L(-(r_c^2+r^2(x_2-y)(x_3+y-1)))$\\$+rx_1r_c\psi^t)\Psi_4+2(x_3+y-1)(r_c\psi^t+r\psi^L(x_2-y))\Psi_2)]\Phi^A$\\$ + [2(r(x_3+y-1)(\psi^L(x_2'-r^2x_2'+r^2y)$\\$-rr_c\psi^t)\Psi_3^{-+}+(r\psi^L(r_c^2+(x_2-y)(x_3+y-1))$\\$+r_c\psi^t(r^2(x_3+y-1)+x_2-y))\Psi_3^{+-})]\Phi^T$}\\
$H_{T_{d1}}^{LL}$   &\tabincell{c}{$[2(r^2-1)^2r_cx_3'\psi^t\Psi_4]\Phi^A$}
                    &\tabincell{c}{$[4(r^2-1)x_3'(r\psi^L(x_1-y)-r_c\psi^t)\Psi_4]\Phi^A$\\$ + [2r(x_1-1)(\psi^L((r^2-1)x_1'-r^2y$\\$+1)-rr_c\psi^t)(\Psi_3^{-+}+\Psi_3^{+-})]\Phi^T$}\\
$H_{T_{d1}}^{SP}$   &\tabincell{c}{$[2r(r^2-1)^2(x_1-y)x_3'\psi^L\Psi_4]\Phi^A$\\$ + [(r^2-1)(1-x_1)(r\psi^L(x_1-y)-r_c\psi^t)(\Psi_3^{-+}+\Psi_3^{+-})]\Phi^T$}
                    &\tabincell{c}{$[2(1-r^2)(x_1-1)(r\psi^L(x_1-y)-r_c\psi^t)(\Psi_3^{-+}+\Psi_3^{+-})]\Phi^T$}\\
$H_{T_{d2}}^{LL}$   &\tabincell{c}{$[2r(1-r^2)x_3((r^2-1)x_1'-r^2y+1)\psi^L\Psi_4]\Phi^A$}
                    &$[4rx_3(\psi^L((r^2-1)x_1'-r^2y+1)-rr_c\psi^t)\Psi_4]\Phi^A$\\
$H_{T_{d2}}^{SP}$   &\tabincell{c}{$[2(1-r^2)r^2x_3r_c\psi^t\Psi_4]\Phi^A$}
                    &$0$\\
$H_{T_{d6}}^{LL}$   &\tabincell{c}{$[(r(r^2-1)\psi^L(-r_c^2-(1-x_2-y)(x_1'$\\$-r^2(x_1'-y)-1))+(r^2-1)r_c\psi^tx_3')\Psi_4]\Phi^V$\\$ + [(r^2-1)(r\psi^L((x_2+y-1)(r^2(+y-x_1')-x_2'-x_3')-r_c^2)$\\$+(r^2-1)r_c\psi^tx_3')\Psi_4-2(r^2-1)r_c\psi^t(x_2+y-1)\Psi_2]\Phi^A$}
                    &\tabincell{c}{$[-2(r\psi^L(-2r_c^2+r^2((x_3+2y-2)x_2'+(y-1)(x_3'+x_3$\\$+2y-2)+x_2(x_2'-x_1'+2y-1))-(x_2-x_1+2y-1)x_2'$\\$-(x_2+y)x_3'+x_3')+r_c\psi^t(r^2(x_3'+x_3)-x_3'))\Psi_4]\Phi^V$\\$ + [2(r_c\psi^t(r^2(x_3-x_3')+x_3')+r\psi^L(r^2(x_3(x_2'+y-1)$\\$-(x_2+y-1)x_3')+(x_2+y-1)x_3'-x_3x_2'))\Psi_4$\\$+4(x_2+y-1)(r_c\psi^t+r\psi^L(y-x_1))\Psi_2]\Phi^A$\\$ + [2r(r_c^2\psi^L(\Psi_3^{-+}+\Psi_3^{+-})-rr_c\psi^t(x_1+x_3-y)\Psi_3^{-+}$\\$+\psi^L(x_1+x_3-y)(r^2y+x_1'(1-r^2)-1)\Psi_3^{+-})]\Phi^T$}\\
$H_{T_{d6}}^{SP}$   &\tabincell{c}{$[r(r^2-1)(\psi^L(-r_c^2-(x_1-y)(r^2(y-x_3')$\\$-r^2x_1'-x_2'))+rx_3r_c\psi^t)\Psi_4]\Phi^V$\\$ + [r(r^2-1)((\psi^L((y-x_1)(r^2(x_2'+y-1)-x_2')-r_c^2)$\\$+rx_3r_c\psi^t)\Psi_4+2(r^2-1)(x_2+y-1)(y-x_1)\psi^L\Psi_2)]\Phi^A$\\$ + [(1-r^2)((x_1-y)(r\psi^L(x_1+x_3-y)$\\$-r_c\psi^t)\Psi_3^{-+}+(r^2-1)x_3r_c\psi^t\Psi_3^{+-})]\Phi^T$}
                    &\tabincell{c}{$[2(1-r^2)((x_1-y)(r\psi^L(1-x_2$\\$-y)-r_c\psi^t)\Psi_3^{-+}+x_3r_c\psi^t\Psi_3^{+-})]\Phi^T$}\\
\hline\hline
\end{tabular}
\end{table}

\end{appendix}


\begin{thebibliography}{99}

\bibitem{Neubert:2001sj}
M.~Neubert and A.~A.~Petrov,
Comments on color suppressed hadronic $B$ decays,
Phys. Lett. B \textbf{519}, 50 (2001).

\bibitem{CDF:1995izg}
F.~Abe \textit{et al.} (CDF Collaboration),
Reconstruction of $B^0 \rightarrow J/\psi K_s^0$ and Measurement of Ratios of Branching Ratios Involving $B \rightarrow J/\psi K^*$,
Phys. Rev. Lett. \textbf{76}, 2015  (1996).

\bibitem{Belle:2002oex}
K.~Abe \textit{et al.} (Belle Collaboration),
Measurement of branching fractions and charge asymmetries for two-body $B$ meson decays with charmonium,
Phys. Rev. D \textbf{67}, 032003 (2003).

\bibitem{BaBar:2004htr}
B.~Aubert \textit{et al.} (BaBar Collaboration),
Measurement of Branching Fractions and Charge Asymmetries for Exclusive $B$ Decays to Charmonium,
Phys. Rev. Lett. \textbf{94}, 141801 (2005).

\bibitem{prd71114008}
C.~H.~Chen and H.~N.~Li,
Nonfactorizable contributions to $B$ meson decays into charmonia,
Phys. Rev. D \textbf{71}, 114008 (2005).

\bibitem{Cheng:2000kt}
H.~Y.~Cheng and K.~C.~Yang,
$B \rightarrow J / \psi K$ decays in QCD factorization,
Phys. Rev. D \textbf{63}, 074011 (2001).

\bibitem{Buchalla:1995vs}
G.~Buchalla, A.~J.~Buras, and M.~E.~Lautenbacher,
Weak decays beyond leading logarithms,
Rev. Mod. Phys. \textbf{68}, 1125 (1996).

\bibitem{Mannel:1997pc}
T.~Mannel and S.~Recksiegel,
Probing the helicity structure of $b \rightarrow s \gamma$ in $\Lambda_b \rightarrow \Lambda \gamma$,
Acta Phys. Pol. B \textbf{28}, 2489 (1997).

\bibitem{Hiller:2007ur}
G.~Hiller, M.~Knecht, F.~Legger, and T.~Schietinger,
Photon polarization from helicity suppression in radiative decays of polarized $\Lambda_b$ to spin 3/2 baryons,
Phys. Lett. B \textbf{649}, 152 (2007).

\bibitem{UA1:1991vse}
C.~Albajar \textit{et al.} (UA1 Collaboration),
First observation of the beauty baryon $\Lambda_b$ in the decay channel $\Lambda_b\rightarrow J/\psi \Lambda$ at the CERN proton-anti-proton collider,
Phys. Lett. B \textbf{273}, 540 (1991).

\bibitem{CDF:1992lrw}
F.~Abe \textit{et al.} (CDF Collaboration),
Search for $\Lambda_b \rightarrow J/\psi \Lambda^0$ in $p\bar{p}$ collisions at $\sqrt{s} = 1.8$ TeV,
Phys. Rev. D \textbf{47}, R2639 (1993).

\bibitem{CDF:2006eul}
A.~Abulencia \textit{et al.} (CDF Collaboration),
Measurement of the $\Lambda^0_{b}$ Lifetime in $\Lambda^0_{b} \rightarrow J/\psi \Lambda^0$ in $p \bar{p}$ Collisions at $\sqrt{s}$ = 1.96 TeV,
Phys. Rev. Lett. \textbf{98}, 122001 (2007).

\bibitem{CDF:1996rvy}
F.~Abe \textit{et al.} (CDF Collaboration),
Observation of $\Lambda_b^0 \rightarrow J/\psi \Lambda$ at the Fermilab proton antiproton collider,
Phys. Rev. D \textbf{55}, 1142 (1997).

\bibitem{D0:2004quf}
V.~M.~Abazov \textit{et al.} (D0 Collaboration),
Measurement of the $\Lambda_b^0$ Lifetime in the Decay $\Lambda_b^0 \rightarrow J/\psi \Lambda^0$ with the D0 Detector,
Phys. Rev. Lett. \textbf{94}, 102001 (2005).

\bibitem{D0:2007giz}
V.~M.~Abazov \textit{et al.} (D0 Collaboration),
Measurement of the $\Lambda_b$ Lifetime in the Exclusive Decay $\Lambda_b \rightarrow J/\psi \Lambda$,
Phys. Rev. Lett. \textbf{99}, 142001 (2007).

\bibitem{prd84031102}
V.~M.~Abazov \textit{et al.} (D0 Collaboration),
Measurement of the production fraction times branching fraction $f(b\to\Lambda_{b})\cdot \mathcal{B}(\Lambda_{b}\to J/\psi \Lambda)$,
Phys. Rev. D \textbf{84}, 031102 (2011).

\bibitem{prd85112003}
V.~M.~Abazov \textit{et al.} (D0 Collaboration),
Measurement of the $\Lambda_b^0$ lifetime in the exclusive decay $\Lambda_b^0 \to J/\psi \Lambda^0$ in $p\bar{p}$ collisions at $\sqrt{s}=1.96$ TeV,
Phys. Rev. D \textbf{85}, 112003 (2012).

\bibitem{pdg2020}
Particle Data Group,
Review of particle physics,
Prog. Theor. Exp. Phys. \textbf{2020}, 083C01 (2020).

\bibitem{plb72427}
R.~Aaij \textit{et al.} (LHCb Collaboration),
Measurements of the $\Lambda_b^0 \to J/\psi \Lambda$ decay amplitudes and the $\Lambda_b^0$ polarisation in $pp$ collisions at $\sqrt{s} = 7$ TeV,
Phys. Lett. B \textbf{724}, 27 (2013).

\bibitem{prd89092009}
G.~Aad \textit{et al.} (ATLAS Collaboration),
Measurement of the parity-violating asymmetry parameter $\alpha_b$ and the helicity amplitudes for the decay $\Lambda_b^0\to J/\psi+\Lambda^0$ with the ATLAS detector,
Phys. Rev. D \textbf{89}, 092009 (2014).

\bibitem{CMS:2016iaf}
(CMS Collaboration),
Measurement of $\Lambda_b$  polarization and the angular parameters of the decay $\Lambda_b \rightarrow \Lambda J/\psi$,
Report No. CMS-PAS-BPH-15-002, 2016.

\bibitem{prd97072010}
A.~M.~Sirunyan \textit{et al.} (CMS Collaboration),
Measurement of the $\Lambda_b$ polarization and angular parameters in $\Lambda_b\to J/\psi\, \Lambda$ decays from $pp$ collisions at $\sqrt{s}=$ 7 and 8 TeV,
Phys. Rev. D \textbf{97}, 072010 (2018).

\bibitem{jhep062020110}
R.~Aaij \textit{et al.} (LHCb Collaboration),
Measurement of the $\Lambda^0_b\rightarrow J/\psi\Lambda$ angular distribution and the $\Lambda^0_b$ polarisation in $pp$ collisions,
J. High Energy Phys. 06(2020)110.

\bibitem{LHCb:2016hey}
R.~Aaij \textit{et al.} (LHCb Collaboration),
Observation of $ \Lambda_b^0 \to \psi(2S)pK^-$ and $ \Lambda_b^0 \to J/\psi \pi^+ \pi^- pK^-$ decays and a measurement of the $\Lambda_b^0$ baryon mass,
J. High Energy Phys. 05(2016)132.

\bibitem{LHCb:2017zzt}
R.~Aaij \textit{et al.} (LHCb Collaboration),
Observation of the Decays $\Lambda_b^0 \to \chi_{c1} p K^-$ and $\Lambda_b^0 \to \chi_{c2} p K^-$,
Phys. Rev. Lett. \textbf{119}, 062001 (2017).

\bibitem{LHCb:2018qxv}
R.~Aaij \textit{et al.} (LHCb Collaboration),
Observation of the decay $\Lambda^0_b\rightarrow\psi(2S)p\pi^-$,
J. High Energy Phys. 08(2018)131.

\bibitem{LHCb:2019imv}
R.~Aaij \textit{et al.} (LHCb Collaboration),
Observation of the $\Lambda_b^0\rightarrow \chi_{c1}(3872)pK^-$ decay,
J. High Energy Phys. 09(2019)028.

\bibitem{plb75163}
G.~Aad \textit{et al.} (ATLAS Collaboration),
Measurement of the branching ratio $\Gamma(\Lambda_b^0 \rightarrow \psi(2S)\Lambda^0)/\Gamma(\Lambda_b^0 \rightarrow J/\psi\Lambda^0)$ with the ATLAS detector,
Phys. Lett. B \textbf{751}, 63 (2015).

\bibitem{jhep032019126}
R.~Aaij \textit{et al.} (LHCb Collaboration),
Measurement of the ratio of branching fractions of the decays $\Lambda^0_b\to\psi(2S) \Lambda$ and $\Lambda^0_b\!\to J/\psi \Lambda$,
J. High Energy Phys. 03(2019)126.

\bibitem{Leibovich:2003tw}
A.~K.~Leibovich, Z.~Ligeti, I.~W.~Stewart, and M.~B.~Wise,
Predictions for nonleptonic $\Lambda_b$ and $\Theta_b$ decays,
Phys. Lett. B \textbf{586}, 337 (2004).

\bibitem{Mohanta:1998iu}
R.~Mohanta, A.~K.~Giri, M.~P.~Khanna, M.~Ishida, S.~Ishida, and M.~Oda,
Hadronic weak decays of $\Lambda_b$ baryon in the covariant oscillator quark model,
Prog. Theor. Phys. \textbf{101}, 959 (1999).

\bibitem{prd562799}
H.~Y.~Cheng,
Nonleptonic weak decays of bottom baryons,
Phys. Rev. D \textbf{56}, 2799 (1997) [\textbf{99}, 079901(E) (2019)].

\bibitem{prd58014016}
Fayyazuddin and Riazuddin,
Two-body nonleptonic $\Lambda_b$ decays in quark model with factorization ansatz,
Phys. Rev. D \textbf{58}, 014016 (1998).

\bibitem{mpla13181}
M.~A.~Ivanov, J.~G.~Korner, V.~E.~Lyubovitskij, and A.~G.~Rusetsky,
Exclusive nonleptonic bottom to charm baryon decays including nonfactorizable contributions,
Mod. Phys. Lett. A \textbf{13}, 181 (1998).

\bibitem{prd80094016}
Z.~T.~Wei, H.~W.~Ke, and X.~Q.~Li,
Evaluating decay rates and asymmetries of $\Lambda_b$ into light baryons in LFQM,
Phys. Rev. D \textbf{80}, 094016 (2009).

\bibitem{ijmpa271250016}
L.~Mott and W.~Roberts,
Rare dileptonic decays of $\Lambda_b$ in a quark model,
Int. J. Mod. Phys. A \textbf{27}, 1250016 (2012).

\bibitem{prd95113002}
Fayyazuddin and M.~J.~Aslam,
Hadronic weak decay $\mathcal{B}_{b}(\frac{1}{2}^+) \to \mathcal{B}(\frac{1}{2}^{+},\; \frac{3}{2}^{+}) +V$,
Phys. Rev. D \textbf{95}, 113002 (2017).

\bibitem{prd65074030}
C.~H.~Chou, H.~H.~Shih, S.~C.~Lee, and H.~n.~Li,
$\Lambda_b\rightarrow \Lambda J/\psi$ decay in perturbative QCD,
Phys. Rev. D \textbf{65}, 074030 (2002).

\bibitem{prd99054020}
J.~Zhu, Z.~T.~Wei, and H.~W.~Ke,
Semileptonic and nonleptonic weak decays of $\Lambda_b^0$,
Phys. Rev. D \textbf{99}, 054020 (2019).

\bibitem{prd575632}
M.~A.~Ivanov, J.~G.~Korner, V.~E.~Lyubovitskij, and A.~G.~Rusetsky,
Exclusive nonleptonic decays of bottom and charm baryons in a relativistic three quark model: Evaluation of nonfactorizing diagrams,
Phys. Rev. D \textbf{57}, 5632 (1998).

\bibitem{prd531457}
H.~Y.~Cheng and B.~Tseng,
1/M corrections to baryonic form-factors in the quark model,
Phys. Rev. D \textbf{53}, 1457 (1996)[\textbf{55}, 1697(E) (1997)].

\bibitem{prd92114013}
Y.~K.~Hsiao, P.~Y.~Lin, C.~C.~Lih, and C.~Q.~Geng,
Charmful two-body anti-triplet $b$-baryon decays,
Phys. Rev. D \textbf{92}, 114013 (2015).

\bibitem{prd96013003}
T.~Gutsche, M.~A.~Ivanov, J.~G.~K\"orner, V.~E.~Lyubovitskij, V.~V.~Lyubushkin, and P.~Santorelli,
Theoretical description of the decays $\Lambda_b \to \Lambda^{(\ast)}(\frac12^\pm,\frac32^\pm) + J/\psi$,
Phys. Rev. D \textbf{96}, 013003 (2017).

\bibitem{prd88114018}
T.~Gutsche, M.~A.~Ivanov, J.~G.~K\"orner, V.~E.~Lyubovitskij, and P.~Santorelli,
Polarization effects in the cascade decay in the covariant confined quark model,
Phys. Rev. D \textbf{88}, 114018 (2013).

\bibitem{prd92114008}
T.~Gutsche, M.~A.~Ivanov, J.~G.~K\"orner, V.~E.~Lyubovitskij, and P.~Santorelli,
Towards an assessment of the ATLAS data on the branching ratio,
Phys. Rev. D \textbf{92}, 114008 (2015).

\bibitem{plb614165}
Z.~J.~Ajaltouni, E.~Conte, and O.~Leitner,
$\Lambda_b$ decays into $\Lambda$-vector,
Phys. Lett. B \textbf{614}, 165 (2005).

\bibitem{plb751127}
Y.~K.~Hsiao, P.~Y.~Lin, L.~W.~Luo, and C.~Q.~Geng,
Fragmentation fractions of two-body $b$-baryon decays,
Phys. Lett. B \textbf{751}, 127 (2015).

\bibitem{prd98074011}
T.~Gutsche, M.~A.~Ivanov, J.~G.~K\"orner, and V.~E.~Lyubovitskij,
Nonleptonic two-body decays of single heavy baryons $\Lambda_Q$, $\Xi_Q$, and $\Omega_Q$ $(Q=b,c)$ induced by $W$ emission in the covariant confined quark model,
Phys. Rev. D \textbf{98}, 074011 (2018).

\bibitem{Xing:2022uqu}
Z.~P.~Xing, F.~Huang, and W.~Wang,
Angular distributions for $\Lambda_b\to \Lambda^*_J (pK^-)J/\psi$ decays,
arXiv:/2203.13524.

\bibitem{prd59094014}
H.~H.~Shih, S.~C.~Lee, and H.~n.~Li,
The $\Lambda_b \rightarrow p l \bar{\nu}$ decay in perturbative QCD,
Phys. Rev. D \textbf{59}, 094014 (1999).

\bibitem{prd61114002}
H.~H.~Shih, S.~C.~Lee, and H.~n.~Li,
Applicability of perturbative QCD to  $\Lambda_b \rightarrow \Lambda_c$  decays,
Phys. Rev. D \textbf{61}, 114002 (2000).

\bibitem{cjp39328}
H.~H.~Shih, S.~C.~Lee, and H.~N.~Li,
Asymmetry parameter in the polarized $\Lambda_b \rightarrow \Lambda_c l \bar{\nu}$ decay,
Chin. J. Phys. \textbf{39}, 328 (2001).

\bibitem{prd74034026}
X.~G.~He, T.~Li, X.~Q.~Li, and Y.~M.~Wang,
PQCD calculation for $\Lambda_b \rightarrow \Lambda \gamma$ in the standard model,
Phys. Rev. D \textbf{74}, 034026 (2006).

\bibitem{prd80034011}
C.~D.~Lu, Y.~M.~Wang, H.~Zou, A.~Ali, and G.~Kramer,
Anatomy of the pQCD approach to the baryonic decays $\Lambda_b\rightarrow p\pi, pK$,
Phys. Rev. D \textbf{80}, 034011 (2009).

\bibitem{220204804}
J.~J.~Han, Y.~Li, H.~n.~Li, Y.~L.~Shen, Z.~J.~Xiao, and F.~S.~Yu,
$\Lambda_b\to p$ transition form factors in perturbative QCD,
Eur. Phys. J. C \textbf{82}, 686 (2022).

\bibitem{220209181}
C.~Q.~Zhang, J.~M.~Li, M.~K.~Jia, and Z.~Rui,
Nonleptonic two-body decays of $\Lambda_b\rightarrow \Lambda_c \pi, \Lambda_cK$ in the perturbative QCD approach,
Phys. Rev. D \textbf{105}, 073005 (2022).

\bibitem{Ali:2012zza}
A.~Ali, C.~Hambrock, and A.~Y.~Parkhomenko,
Light-cone wave functions of heavy baryons,
Theor. Math. Phys. \textbf{170}, 2 (2012).

\bibitem{plb665197}
P.~Ball, V.~M.~Braun, and E.~Gardi,
Distribution amplitudes of the $\Lambda_b$ baryon in QCD,
Phys. Lett. B \textbf{665}, 197 (2008).

\bibitem{jhep112013191}
G.~Bell, T.~Feldmann, Y.~M.~Wang, and M.~W.~Y.~Yip,
Light-cone distribution amplitudes for heavy-quark hadrons,
J. High Energy Phys. 11(2013)191.

\bibitem{jhep022016179}
Y.~M.~Wang and Y.~L.~Shen,
Perturbative corrections to $\Lambda_b \rightarrow \Lambda$ form factors from QCD Light-cone sum rules,
J. High Energy Phys. 02(2016)179.

\bibitem{zpc42569}
V.~L.~Chernyak, A.~A.~Ogloblin, and I.~R.~Zhitnitsky,
Wave functions of octet baryons,
Z. Phys. C \textbf{42}, 569 (1989).

\bibitem{prd90114030}
Z.~Rui and Z.~T.~Zou,
$S$-wave ground state charmonium decays of $B_c$ mesons in the perturbative QCD approach,
Phys. Rev. D \textbf{90}, 114030 (2014).

\bibitem{epjc75293}
Z.~Rui, W.~F.~Wang, G.~x.~Wang, L.~h.~Song, and C.~D.~L\"u,
The $B_c\rightarrow \psi (2S)\pi $ , $\eta _c(2S)\pi $ decays in the perturbative QCD approach,
Eur. Phys. J. C \textbf{75}, 293 (2015).

\bibitem{epjc76564}
Z.~Rui, H.~Li, G.~x.~Wang, and Y.~Xiao,
Semileptonic decays of $B_c$ meson to $S$-wave charmonium states in the perturbative QCD approach,
Eur. Phys. J. C \textbf{76}, 564 (2016).

\bibitem{epjc77199}
Z.~Rui, Y.~Li, and W.~F.~Wang,
The S-wave resonance contributions in the $B^0_s$ decays into $ \psi(2S,3S)$ plus pion pair,
Eur. Phys. J. C \textbf{77}, 199 (2017).

\bibitem{epjc77610}
Z.~Rui, Y.~Li and, Z.~J.~Xiao,
Branching ratios, $CP$ asymmetries and polarizations of $B\rightarrow \psi(2S) V$ decays,
Eur. Phys. J. C \textbf{77}, 610 (2017).

\bibitem{prd97033006}
Z.~Rui and W.~F.~Wang,
$S$-wave $K\pi$ contributions to the hadronic charmonium $B$ decays in the perturbative QCD approach,
Phys. Rev. D \textbf{97}, 033006 (2018).

\bibitem{prd98113003}
Z.~Rui, Y.~Li, and H.~N.~Li,
$P$-wave contributions to $B \rightarrow \psi\pi\pi$ decays in perturbative QCD approach,
Phys. Rev. D \textbf{98},  113003 (2018).

\bibitem{prd99093007}
Z.~Rui, Y.~Q.~Li, and J.~Zhang,
Isovector scalar $a_0(980)$ and $a_0(1450)$ resonances in the $B\rightarrow \psi (K\bar{K},\pi\eta)$ decays,
Phys. Rev. D \textbf{99}, 093007 (2019).

\bibitem{epjc79792}
Z.~Rui, Y.~Li, and H.~Li,
Studies of the resonance components in the $B_s$ decays into charmonia plus kaon pair,
Eur. Phys. J. C \textbf{79}, 792 (2019).

\bibitem{cpc44073102}
Y.~Li, Z.~Rui, and Z.~J.~Xiao,
$P$-wave contributions to $B_{(s)} \rightarrow \psi K\pi$ decays in perturbative QCD approach,
Chin. Phys. C \textbf{44}, 073102 (2020).

\bibitem{prd101016015}
Y.~Li, D.~C.~Yan, Z.~Rui, and Z.~J.~Xiao,
$S$, $P$ and $D$-wave resonance contributions to $B_{(s)} \rightarrow \eta_c(1S,2S) K\pi$ decays in the perturbative QCD approach,
Phys. Rev. D \textbf{101}, 016015 (2020).

\bibitem{Liu:2018kuo}
X.~Liu, H.~n.~Li, and Z.~J.~Xiao,
Improved perturbative QCD formalism for $B_c$ meson decays,
Phys. Rev. D \textbf{97}, 113001 (2018).

\bibitem{Liu:2020upy}
X.~Liu, H.~n.~Li, and Z.~J.~Xiao,
Next-to-leading-logarithm $k_T$ resummation for $B_c\to J/\psi$ decays,
Phys. Lett. B \textbf{811}, 135892 (2020).

\bibitem{Beneke:1999br}
M.~Beneke, G.~Buchalla, M.~Neubert, and C.~T.~Sachrajda,
QCD Factorization for $B \rightarrow \pi \pi$ Decays: Strong Phases and CP Violation in the Heavy Quark Limit,
Phys. Rev. Lett. \textbf{83}, 1914 (1999).

\bibitem{Beneke:2000ry}
M.~Beneke, G.~Buchalla, M.~Neubert, and C.~T.~Sachrajda,
QCD factorization for exclusive, nonleptonic $B$ meson decays: General arguments and the case of heavy light final states,
Nucl. Phys. \textbf{B591}, 313 (2000).

\bibitem{Beneke:2003zv}
M.~Beneke and M.~Neubert,
QCD factorization for $B \rightarrow PP$ and $B \rightarrow PV$ decays,
Nucl. Phys. \textbf{B675}, 333 (2003).

\bibitem{epjc732302}
A.~Ali, C.~Hambrock, A.~Y.~Parkhomenko, and W.~Wang,
Light-cone distribution amplitudes of the ground state bottom baryons in HQET,
Eur. Phys. J. C \textbf{73}, 2302 (2013).

\bibitem{plb738334}
V.~M.~Braun, S.~E.~Derkachov, and A.~N.~Manashov,
Integrability of the evolution equations for heavy-light baryon distribution amplitudes,
Phys. Lett. B \textbf{738}, 334 (2014).

\bibitem{Groote:1997yr}
S.~Groote, J.~G.~Korner, and O.~I.~Yakovlev,
An analysis of diagonal and nondiagonal QCD sum rules for heavy baryons at next-to-leading order in $\alpha_s$,
Phys. Rev. D \textbf{56}, 3943 (1997).

\bibitem{220506095}
K.~S.~Huang, W.~Liu, Y.~L.~Shen, and F.~S.~Yu,
$\Lambda_b \rightarrow p, N^\ast(1535)$ form factors from QCD Light-cone sum rules,
arXiv:2205.06095.

\bibitem{Farrar:1988vz}
G.~R.~Farrar, H.~Zhang, A.~A.~Ogloblin, and I.~R.~Zhitnitsky,
Baryon wave functions and cross-sections for photon annihilation to baryon pairs,
Nucl. Phys. \textbf{B311}, 585 (1989).

\bibitem{Liu:2014uha}
Y.~L.~Liu, C.~Y.~Cui, and M.~Q.~Huang,
Higher order light-cone distribution amplitudes of the $\Lambda$ baryon,
Eur. Phys. J. C \textbf{74}, 3041 (2014).

\bibitem{Liu:2008yg}
Y.~L.~Liu and M.~Q.~Huang,
Distribution amplitudes of $\Sigma$ and $\Lambda$ and their electromagnetic form factors,
Nucl. Phys. \textbf{A821}, 80 (2009).

\bibitem{jhep020702016}
G.~S.~Bali, V.~M.~Braun, M.~G\"ockeler, M.~Gruber, F.~Hutzler, A.~Sch\"afer, R.~W.~Schiel, J.~Simeth, W.~S\"oldner, A.~Sternbeck \textit{et al.},
Light-cone distribution amplitudes of the baryon octet,
J. High Energy Phys. 02 (2016) 070.

\bibitem{prd89094511}
V.~M.~Braun, S.~Collins, B.~Gl\"a\ss{}le, M.~G\"ockeler, A.~Sch\"afer, R.~W.~Schiel, W.~S\"oldner, A.~Sternbeck, and P.~Wein,
Light-cone distribution amplitudes of the nucleon and negative parity nucleon resonances from lattice QCD,
Phys. Rev. D \textbf{89}, 094511 (2014).

\bibitem{epja55116}
G.~S.~Bali \textit{et al.} (RQCD Collaboration),
Light-cone distribution amplitudes of octet baryons from lattice QCD,
Eur. Phys. J. A \textbf{55}, 116 (2019).

\bibitem{Bodwin:1994jh}
G.~T.~Bodwin, E.~Braaten, and G.~P.~Lepage,
Rigorous QCD analysis of inclusive annihilation and production of heavy quarkonium,
Phys. Rev. D \textbf{51}, 1125 (1995).

\bibitem{Beneke:1998ks}
M.~Beneke, F.~Maltoni, and I.~Z.~Rothstein,
QCD analysis of inclusive $B$ decay into charmonium,
Phys. Rev. D \textbf{59}, 054003 (1999).

\bibitem{Jia:2008ep}
Y.~Jia and D.~Yang,
Refactorizing NRQCD short-distance coefficients in exclusive quarkonium production,
Nucl. Phys. \textbf{B814}, 217 (2009).

\bibitem{Chen:2011pn}
Y.~C.~Chen and H.~n.~Li,
Three-parton contribution to pion form factor in $k_T$ factorization,
Phys. Rev. D \textbf{84}, 034018 (2011).

\bibitem{Chen:2011gv}
Y.~C.~Chen and H.~N.~Li,
Three-parton contribution to the $B\to\pi$ form factors in $k_T$ factorization,
Phys. Lett. B \textbf{712}, 63 (2012).

\bibitem{Belyaev:1994zk}
V.~M.~Belyaev, V.~M.~Braun, A.~Khodjamirian, and R.~Ruckl,
$D^* D \pi$ and $B^* B \pi$ couplings in QCD,
Phys. Rev. D \textbf{51}, 6177 (1995).

\bibitem{zpc55659}
J.~G.~Korner and M.~Kramer,
Exclusive nonleptonic charm baryon decays,
Z. Phys. C \textbf{55}, 659 (1992).

\bibitem{Botts:1989kf}
J.~Botts and G.~F.~Sterman,
Hard elastic scattering in QCD: Leading behavior,
Nucl. Phys. \textbf{B325}, 62 (1989).

\bibitem{Kundu:1998gv}
B.~Kundu, H.~n.~Li, J.~Samuelsson, and P.~Jain,
The perturbative proton form-factor reexamined,
Eur. Phys. J. C \textbf{8}, 637 (1999).

\bibitem{prd102054511}
D.~Hatton \textit{et al.} (HPQCD Collaboration),
Charmonium properties from lattice QCD+QED: Hyperfine splitting, $J/\psi$ leptonic width, charm quark mass, and $a^c_\mu$,
Phys. Rev. D \textbf{102}, 054511 (2020).

\bibitem{jhep120652020}
H.~S.~Chung,
$\overline {MS}$ renormalization of $S$-wave quarkonium wavefunctions at the origin,
J. High Energy Phys. 12(2020)065.

\bibitem{Gutsche:2013pp}
T.~Gutsche, M.~A.~Ivanov, J.~G.~Korner, V.~E.~Lyubovitskij, and P.~Santorelli,
Rare baryon decays $\Lambda_b \to \Lambda {l^{+}l^{-}} (l=e, \mu, \tau)$ and $\Lambda_b \to \Lambda\gamma$: Differential and total rates, lepton- and hadron-side forward-backward asymmetries,
Phys. Rev. D \textbf{87}, 074031 (2013).

\bibitem{Mohanta:2000nk}
R.~Mohanta, A.~K.~Giri, and M.~P.~Khanna,
Charmless two-body hadronic decays of $\Lambda_b$ baryon,
Phys. Rev. D \textbf{63}, 074001 (2001).

\bibitem{Detmold:2016pkz}
W.~Detmold and S.~Meinel,
$\Lambda_b \rightarrow \Lambda \ell^+ \ell^-$ form factors, differential branching fraction, and angular observables from lattice QCD with relativistic $b$ quarks,
Phys. Rev. D \textbf{93}, 074501 (2016).

\bibitem{Aliev:2010uy}
T.~M.~Aliev, K.~Azizi, and M.~Savci,
Analysis of the $\Lambda_{b}\rightarrow \Lambda \ell^+\ell^- $ decay in QCD,
Phys. Rev. D \textbf{81}, 056006 (2010).

\bibitem{Huang:1998ek}
C.~S.~Huang and H.~G.~Yan,
Exclusive rare decays of heavy baryons to light baryons: $\Lambda(b) \rightarrow \Lambda \gamma$ and $\Lambda_b \rightarrow \Lambda l^+ l^-$,
Phys. Rev. D \textbf{59}, 114022 (1999)[\textbf{61}, 039901(E) (2000)].

\bibitem{prd89094010}
X.~Liu, W.~Wang, and Y.~Xie,
Penguin pollution in $B \rightarrow J/\psi V$ decays and impact on the extraction of the $B_s-\bar B_s$ mixing phase,
Phys. Rev. D \textbf{89}, 094010 (2014).

\bibitem{Astbury:1975hn}
P.~Astbury, J.~Gallivan, J.~Jafar, M.~Letheren, V.~Steiner, J.~A.~Wilson, W.~Beusch, M.~Borghini, D.~Websdale, L.~Fluri \textit{et al.},
Measurement of the differential cross-section and the spin-correlation parameters $P$,$A$, and $R$ in the backward peak of $\pi^- p \rightarrow K^0 \Lambda$ at 5 GeV/c,
Nucl. Phys. \textbf{B99}, 30 (1975).

\bibitem{Cleland:1972fa}
W.~E.~Cleland, G.~Conforto, G.~H.~Eaton, H.~J.~Gerber, M.~Reinharz, A.~Gautschi, E.~Heer, C.~Revillard, and G.~Von Dardel,
A measurement of the $\beta$-parameter in the charged nonleptonic decay of the $\Lambda^0$ hyperon,
Nucl. Phys. \textbf{B40}, 221 (1972).

\bibitem{Dauber:1969hg}
P.~M.~Dauber, J.~P.~Berge, J.~R.~Hubbard, D.~W.~Merrill, and R.~A.~Muller,
Production and decay of cascade hyperons,
Phys. Rev. \textbf{179}, 1262 (1969).

\bibitem{Overseth:1967zz}
O.~E.~Overseth and R.~F.~Roth,
Time Reversal Invariance in $\Lambda^0$ Decay,
Phys. Rev. Lett. \textbf{19}, 391 (1967).

\bibitem{Cronin:1963zb}
J.~W.~Cronin and O.~E.~Overseth,
Measurement of the decay parameters of the $\Lambda^0$ particle,
Phys. Rev. \textbf{129}, 1795 (1963).

\bibitem{BESIII:2018cnd}
M.~Ablikim \textit{et al.} (BESIII Collaboration),
Polarization and entanglement in baryon-antibaryon pair production in electron-positron annihilation,
Nat. Phys. \textbf{15}, 631 (2019).

\bibitem{Ireland:2019uja}
D.~G.~Ireland, M.~D\"oring, D.~I.~Glazier, J.~Haidenbauer, M.~Mai, R.~Murray-Smith, and D.~R\"onchen,
Kaon Photoproduction and the $\Lambda$ Decay Parameter $\alpha_-$,
Phys. Rev. Lett. \textbf{123}, 182301 (2019).

\bibitem{Jiang:2018iqa}
H.~Y.~Jiang and F.~S.~Yu,
Fragmentation-fraction ratio $f_{\Xi_b}/f_{\Lambda_b}$ in $b$- and $c$-baryon decays,
Eur. Phys. J. C \textbf{78}, 224 (2018).

\bibitem{jhep110672015}
M.~Galanti, A.~Giammanco, Y.~Grossman, Y.~Kats, E.~Stamou, and J.~Zupan,
Heavy baryons as polarimeters at colliders,
J. High Energy Phys. 11(2015)067.

\bibitem{Leitner:2004dt}
O.~Leitner, Z.~J.~Ajaltouni, and E.~Conte,
An angular distribution analysis of $\Lambda_b$ decays,
Nucl. Phys. \textbf{A755}, 435 (2005).

\bibitem{epjc59861}
Y.~m.~Wang, Y.~Li, and C.~D.~Lu,
Rare decays of $\Lambda_b \rightarrow \Lambda + \gamma$ and $\Lambda_b \rightarrow \Lambda + l^+ l^-$ in the light-cone sum rules,
Eur. Phys. J. C \textbf{59}, 861 (2009).

\bibitem{prd63114024}
C.~H.~Chen and C.~Q.~Geng,
Rare $\Lambda_b \rightarrow \Lambda l^+ l^-$ decays with polarized $\Lambda$,
Phys. Rev. D \textbf{63}, 114024 (2001).

\bibitem{prd78114032}
M.~J.~Aslam, Y.~M.~Wang, and C.~D.~Lu,
Exclusive semileptonic decays of $\Lambda_b \rightarrow \Lambda l^+ l^-$ in supersymmetric theories,
Phys. Rev. D \textbf{78}, 114032 (2008).

\bibitem{prd531416}
W.~Loinaz and R.~Akhoury,
Exclusive semileptonic decays of $b$ baryons into protons,
Phys. Rev. D \textbf{53}, 1416 (1996).

\bibitem{zpc51321}
F.~Hussain, J.~G.~Korner, M.~Kramer, and G.~Thompson,
On heavy baryon decay form-factors,
Z. Phys. C \textbf{51}, 321 (1991).

\bibitem{9211255}
F.~Schlumpf,
Relativistic constituent quark model for baryons,
Ph.D. thesis, Zurich, 1992.

\bibitem{Cheng:2001ez}
H.~Y.~Cheng, Y.~Y.~Keum, and K.~C.~Yang,
$B \rightarrow J/\psi K^{*}$ decays in QCD factorization,
Phys. Rev. D \textbf{65}, 094023 (2002).

\end{thebibliography}
\end{document}